\documentclass[aps,prd,twocolumn,preprintnumbers,nofootinbib,10pt]{revtex4-1}%

\pdfoutput=1
\usepackage{amsmath,amsthm,amssymb,multirow,psfrag}
\usepackage{epsfig}
\usepackage{color}

\graphicspath{{./Figures/}}

\begin{document}

\def\lsim{\mathrel{\rlap{\lower4pt\hbox{\hskip1pt$\sim$}}F
  \raise1pt\hbox{$<$}}}
\def\gsim{\mathrel{\rlap{\lower4pt\hbox{\hskip1pt$\sim$}}
  \raise1pt\hbox{$>$}}}
\newcommand{\vev}[1]{ \left\langle {#1} \right\rangle }
\newcommand{\bra}[1]{ \langle {#1} | }
\newcommand{\ket}[1]{ | {#1} \rangle }
\newcommand{\ev}{ {\rm eV} }
\newcommand{\kev}{{\rm keV}}
\newcommand{\mev}{{\rm MeV}}
\newcommand{\tev}{{\rm TeV}}
\newcommand{\mpl}{$M_{Pl}$}
\newcommand{\mw}{$M_{W}$}
\newcommand{\Ft}{F_{T}}
\newcommand{\Zparity}{\mathbb{Z}_2}
\newcommand{\BLambda}{\boldsymbol{\lambda}}
\newcommand{\met}{\;\not\!\!\!{E}_T}
\newcommand{\suRL}{$M_{W}$}

\newcommand{\beq}{\begin{equation}}
\newcommand{\eeq}{\end{equation}}
\newcommand{\bea}{\begin{eqnarray}}
\newcommand{\eea}{\end{eqnarray}}
\newcommand{\nn}{\nonumber \\ }
\newcommand{\gev}{{\mathrm GeV}}
\newcommand{\hc}{\mathrm{h.c.}}
\newcommand{\eps}{\epsilon}

\newcommand{\cO}{{\cal O}}
\newcommand{\cL}{{\cal L}}
\newcommand{\cM}{{\cal M}}

\newcommand{\Deltab}{ {{\bar{\Delta}}} }
\newcommand{\Hb}{ {{\bar{H}}} }
\newcommand{\Phb}{ {{\bar{\Phi}}} }
\newcommand{\Tr}[1]{\ensuremath{{\bf Tr}[\,#1\,]}} 

\newcommand{\Chi}{ \ensuremath{X} }
\newcommand{\Phib}{ \ensuremath{\pmb{\Phi}} }
\newcommand{\Chid}{\Chi^\dagger}
\newcommand{\Phid}{\Phi^\dagger}
\newcommand{\Chic}{\Chi_c}
\newcommand{\Phic}{\Phi_c}
\newcommand{\Chicstar}{\Chi_c^\dagger}
\newcommand{\Phicstar}{\Phi_c^\dagger}
\newcommand{\txx}{\ensuremath{{\Tr{ \Phic \sigma_i \Phi \sigma_j }  \left[ V \Chi^\dagger V^\dagger \right]_{i,j}  + c.c.} }}
\newcommand{\txcO}{\ensuremath{{\Tr{\Chi^\dagger\Chi\Chi^\dagger\Chi} - \Tr{\Chi^\dagger\Chi}^2}}}
\newcommand{\txcTc}{\ensuremath{{\Tr{ \Chid T_i \Chicstar T_j}\, \Tr{ \Phi \sigma_j \Phic\sigma_i} + c.c.}}}
\newcommand{\txcEc}{\ensuremath{{\Tr{ \Chid T_i \Chi T_j }\, \left[ V \Chi V^\dagger \right]_{i,j} + c.c.}}}
\newcommand{\txcFc}{\ensuremath{{\Tr{ \Phid \sigma_i \Phi \sigma_j}\, \left[ V \Chi V^\dagger \right]_{i,j}  + c.c.}}}
\newcommand{\txcIf}{\ensuremath{{\Tr{ \Chid \Chi} \, \Tr{ \Phid \Phi} }}}
\newcommand{\txcIh}{\ensuremath{{ \Tr{ \Chid T_i \Chi}\, \Tr{ \Phid \sigma_i \Phi}}}}
\newcommand{\txcIcO}{\ensuremath{{ \Tr{ \Chicstar  T_i \Chi_c}\, \Tr{ \Phicstar \sigma_i \Phic   } }}}
\newcommand{\txcIc}{\ensuremath{{ \Tr{ \Chid  T_i \Chi T_j }\,  \Tr{ \Phid\sigma_i \Phi \sigma_j }}}}
%
%

\graphicspath{{./Figures/}}

\newcommand{\fref}[1]{Fig.\,\ref{fig:#1}} 
\newcommand{\eref}[1]{Eq.~\eqref{eq:#1}} 
\newcommand{\aref}[1]{Appendix~\ref{app:#1}}
\newcommand{\sref}[1]{Sec.~\ref{sec:#1}}
\newcommand{\tref}[1]{Table~\ref{tab:#1}}  

\def\TY#1{{\bf  \textcolor{red}{[TY: {#1}]}}}
\newcommand{\draftnote}[1]{{\bf\color{blue} #1}}
\newcommand{\draftnoteR}[1]{{\bf\color{red} #1}}

\title{ {\bf \Large{The Supersymmetric Georgi-Machacek Model}\normalsize}}
\author{\bf{Roberto Vega$\,^{a}$,~Roberto Vega-Morales$\,^{b}$,~Keping Xie$\,^{a}$}}

\vspace{-.1cm}
\affiliation{
$^a$Department of Physics, Southern Methodist University, Dallas, TX 75275, USA\\
$^{b}$Departamento de F\'{i}sica Te\'{o}rica y del Cosmos, Universidad de Granada,\\
Campus de Fuentenueva, E-18071 Granada, Spain}

\begin{abstract}

We show that the well known Georgi-Machacek (GM) model can be realized as a limit of the recently constructed Supersymmetric Custodial Higgs Triplet Model (SCTM) which in general contains a significantly more complex scalar spectrum.~We dub this limit of the SCTM, which gives a weakly coupled origin for the GM model at the electroweak scale, the Supersymmetric GM (SGM) model.~We derive a mapping between the SGM and GM models using it to show how a supersymmetric origin implies constraints on the Higgs potential in conventional GM model constructions which would generically not be present.~We then perform a simplified phenomenological study of diphoton and $ZZ$ signals for a pair of benchmark scenarios to illustrate under what circumstances the GM model can mimic the SGM model and when they should be easily distinguishable.
\end{abstract}

\preprint{UG-FT 325/17,~CAFPE 195/17,~SMU-HEP-17-09}

\maketitle

\section{Introduction}\label{sec:Intro}

The discovery of a 125~GeV scalar at the Large Hadron Collider (LHC)~\cite{Aad:2012tfa,Chatrchyan:2012xdj} with Standard Model (SM) Higgs boson like properties~\cite{Falkowski:2013dza} appears to have settled the nature of the electroweak symmetry breaking (EWSB) mechanism.~However, uncertainties in Higgs boson coupling measurements~\cite{Khachatryan:2014kca,Khachatryan:2016vau,Sirunyan:2017exp,Sirunyan:2017tqd,Aaboud:2017oem,Blasi:2017xmc} still leaves room for extended Higgs sectors which contribute non-negligibly to EWSB\,\footnote{We reserve the `Higgs' label for scalars that contribute to EWSB and therefore do not include electroweak singlets.}.~Of course any extended Higgs scalar sector must be carefully constructed in order to satisfy the stringent constraints~\cite{Agashe:2014kda} from electroweak precision data (EWPD).~In particular, measurements of the $\rho$ parameter imply the tree level relation $\rho_{tree} = 1$, which is automatically satisfied by Higgs sectors respecting the well known `custodial' $SU(2)_C$ global symmetry~\cite{Sikivie:1980hm}.

Extended Higgs sectors that include only electroweak doublets with SM like quantum numbers, as in the Minimal Supersymmetric SM (MSSM), automatically preserve custodial symmetry~\cite{Low:2010jp} regardless of whether each doublet obtains the same vacuum expectation value (VEV) or not.~In order to avoid resorting to highly tuned cancelations, larger electroweak representations are constrained by $\rho_{tree} = 1$ to come in ${\bf (N, \bar{N}})$ representations~\cite{Low:2010jp} of the global $SU(2)_L\otimes SU(2)_R$ symmetry which breaks down to the diagonal $SU(2)_C$ subgroup after EWSB.~In contrast to doublets, this requires multiple scalars for a given $SU(2)_L$ representation and furthermore, their VEVs must be `aligned' at tree level.

One of the most thoroughly explored examples of an extended (non-doublet) Higgs sector is the Georgi-Machacek (GM) model~\cite{Georgi:1985nv,Chanowitz:1985ug} which contains a ${\bf (3, \bar{3}})$ in addition to the SM Higgs doublet, which is a ${\bf (2, \bar{2}})$.~The construction of the ${\bf (3, \bar{3}})$ is accomplished by adding \emph{two} electroweak triplets with hypercharges $Y = 1$ and $Y = 0$ whose VEVs are aligned at tree level.~This leads to a rich phenomenology~\cite{Chanowitz:1985ug,Gunion:1989ci,Gunion:1990dt} which has been examined in many recent studies~\cite{Englert:2013zpa,Chiang:2012cn,Chiang:2014bia,Chiang:2015kka,Chiang:2015amq,Degrande:2015xnm,Hartling:2014zca,Hartling:2014aga,Degrande:2015xnm,Logan:2015xpa,Campbell:2016zbp,deFlorian:2016spz,Degrande:2017naf,Logan:2017jpr,Zhang:2017och}.~Though specifying the origin of the new Higgs scalars in the GM model is not necessary for analyzing much its phenomenology, implicitly it is assumed they arise out of a UV sector which explains their presence and ameliorates the fine tuning issues associated with each of their masses as well as the $\rho$ parameter~\cite{Gunion:1990dt}.~Typically it is envisioned that the GM model scalars arise as pseudo Goldstone bosons~\cite{Georgi:1985nv} of a strongly coupled sector whose global symmetry breaking structure~\cite{Bellazzini:2014yua} contains them in its coset\,\footnote{In particular the $SU(5)/SO(5)$ symmetry breaking pattern found in a number of composite Higgs scenarios~\cite{Mrazek:2011iu,Bellazzini:2014yua}, including certain Little Higgs~\cite{Chang:2003un} and the Littlest Higgs Model with $T$-Parity~\cite{Cheng:2003ju,Cheng:2004yc,Low:2004xc}, contains within its coset the same $SU(2)_L\otimes SU(2)_R$ representations as the GM scalar sector.}.

More recently, the Supersymmetric Custodial Higgs Triplet Model (SCTM) was constructed~\cite{Cort:2013foa,Garcia-Pepin:2014yfa,Delgado:2015bwa} in which the Higgs sector contains \emph{three} electroweak triplet chiral superfields, along with the doublets of the MSSM, and a superpotential plus soft SUSY breaking sector which respects the global $SU(2)_L\otimes SU(2)_R$ symmetry.~As a consequence of supersymmetry, this theory inevitably comes along with a `doubling' of the scalar sector (in addition to introducing a new fermion sector) with respect to the original GM model and leads to a significantly more complicated mass spectrum.~However, as we analyze in detail here, in a certain limit one recovers \emph{only} the GM spectrum at low energies.~We dub this limit of the SCTM, the Supersymmetric GM (SGM) model.~In obtaining the GM model Higgs spectrum from an underlying supersymmetric theory we are able to realize a weakly coupled origin for the GM scalar spectrum at the electroweak scale.

In addition to giving a weakly coupled origin for the GM model and, by virtue of being a superymmetric theory, solving the various fine tuning problems of the GM model~\cite{Gunion:1989ci,Gunion:1990dt}, the SGM also inherits other benefits from the SCTM.~As examined in~\cite{Cort:2013foa,Garcia-Pepin:2014yfa}, both tree level and 1-loop effects in the SCTM can contribute the large corrections necessary to explain the observed Higgs mass without needing to resort to heavy stops or stop mixing as needed in the MSSM~\cite{Draper:2011aa}.~It also avoids problems with EWPD which plague non-custodial supersymmetric Higgs triplet models invoked~\cite{Delgado:2012sm} to solve the Higgs mass problem of the MSSM.~The SCTM also allows for a natural connection between the scale of supersymmetry breaking and the scale at which the original global $SU(2)_L\otimes SU(2)_R$ symmetry holds at tree level~\cite{Garcia-Pepin:2014yfa}.~Furthermore, it can be embedded in a gauge mediated supersymmetry breaking framework~\cite{Delgado:2015bwa}.~The custodial symmetry of the SCTM also automatically realizes an `alignment' limit~\cite{Carena:2013ooa,Carena:2014nza,Carena:2015moc} allowing for regions of parameter space which impersonate the SM without decoupling.~In addition, there are possibilities in the SCTM for generating the strong first order phase transition needed for successful electroweak baryogenesis~\cite{Garcia-Pepin:2016hvs}.~Finally, and as we discuss further below, the SGM inherits the potential (neutralino) dark matter candidates of the SCTM~\cite{Delgado:2015aha}. 

In this work we show explicitly how the Higgs scalar spectrum of the GM model arises as a limit of the SCTM and derive a mapping between the Higgs potentials of the SGM and GM models.~We then use this mapping to show how a supersymmetric origin for the GM model implies correlations between operators in the Higgs potential which would otherwise not be present in the conventional GM model.~We also perform a simplified phenomenological study of diphoton and $ZZ$ signals for a pair of benchmark scenarios to illustrate under what circumstances the GM model can mimic the SGM model and when they should be easily distinguishable.~We also discuss other potentially interesting signals as well as ongoing and future directions for further investigation.

\section{Weak Scale GM from the SCTM}\label{sec:GMlimit}

We will define the SGM as the limit of the SCTM in which the scalar spectrum of the conventional GM model is obtained at low energies.~Thus we need to decouple any additional scalars present in the SCTM which are not present in the GM model.~As we will see, this limit corresponds to taking particular soft supersymmetry breaking masses large.~In this section we first briefly review the relevant aspects of the SCTM model Higgs sector before showing how it can be mapped onto the GM model Higgs sector.~We then show explicitly the limit out of which the SGM model arises from the SCTM and show how the mapping between the GM and SGM scalar potentials implies correlations between the quartic couplings in a GM model with a supersymmetric origin.~The custodial fermion superpartner sector is also briefly discussed.

Throughout our analysis we implicitly assume that the scale $\mathcal{M}$ at which the global $SU(2)_L\otimes SU(2)_R$ holds is not too much larger than the electroweak scale $v$ in order to neglect RG evolution effects~\cite{Gunion:1990dt,Garcia-Pepin:2014yfa,Blasi:2017xmc}.~This is an implicit assumption in almost all GM model constructions so that custodial breaking effects due to RG evolution are small and do not invalidate the custodial classification of the Higgs spectrum at the weak scale.~In the SCTM the scale $\mathcal{M}$ can be connected to the supersymmetry breaking scale and in principle much larger~\cite{Garcia-Pepin:2014yfa} than $v$ due to additional VEV `directions' as a consequence of supersymmetry.~However, for present purposes it is sufficient to take $\mathcal{M} \sim B \sim $~TeV and leave a more general analysis including RG and NLO loop effects to future work.

\subsection{Higgs Sector of the SCTM}\label{subsec:SCTM}

The SCTM field content~\cite{Cort:2013foa} possesses, in addition to the two MSSM Higgs electroweak doublet chiral superfields $H_1$ and $H_2$ with hypercharge $\pm 1/2$, three electroweak triplet chiral superfields $\Sigma_0=(\phi^+,\phi^0,\phi^-)^T$, $\Sigma_+=(\psi^{++},\psi^+,\psi^0)^T$, and $\Sigma_{-}=(\chi^0,\chi^-,\chi^{--})^T$, with hypercharges $Y=0,+1,-1$ respectively.~In the $SU(2)_L$ basis the electroweak doublets can be written as,
\bea
\label{eq:H1H2}
H_1 = 
\begin{pmatrix}
H_1^0\\ H_1^-
\end{pmatrix},~
H_2=
\begin{pmatrix}H_2^+\\ H_2^0
\end{pmatrix} ,
\eea
while the three $SU(2)_L$ triplets can be expressed\,\footnote{Note we use a different phase convention as compared to~\cite{Cort:2013foa}.}
as,
\bea
\label{eq:Sigmas}
\Sigma_{-} =
\begin{pmatrix}
\frac{\chi^-}{\sqrt{2}} & -\chi^0\\
\chi^{--}& -\frac{\chi^-}{\sqrt{2}}
\end{pmatrix}&,&~
\Sigma_{+} =
\begin{pmatrix}
\frac{\psi^+}{\sqrt{2}} & -\psi^{++}\\
\psi^{0}& -\frac{\psi^+}{\sqrt{2}}
\end{pmatrix},\nn 
\Sigma_{0} &=&
\begin{pmatrix}
\frac{\phi^0}{\sqrt{2}} & -\phi^+\\
\phi^{-}& -\frac{\phi^0}{\sqrt{2}} 
\end{pmatrix}.~~~~~~~~~~~~~~~~~
\eea
Note that the $Y = 0$ field is complex while the hypercharge $Y = 1$ and $Y = -1$ fields are independent degrees of freedom as compared to the GM model where the $Y = 0$ triplet is real and only one $Y = 1$ triplet is present along with its conjugate.~The difference is of course a consequence of supersymmetry and in particular the requirement of holomorphy of the superpotential and anomaly cancelation.~As emphasized in~\cite{Garcia-Pepin:2014yfa}, this has implications for the vacuum structure of the Higgs potential and the $\rho$ parameter as well as the ratio of the Higgs boson couplings to $WW$ and $ZZ$ pairs, but will not be relevant for present purposes.

These fields can then be organized into bi-doublet and bi-triplet representations of the global $SU(2)_R\times SU(2)_L$,
\bea
\bar H=
\begin{pmatrix} 
H_1\\ H_2 
\end{pmatrix},
~\bar\Delta=
\begin{pmatrix}
-\frac{\Sigma^0}{\sqrt{2}} & \Sigma_{-}\\
-\Sigma_+ & \frac{\Sigma_0}{\sqrt{2}}
\end{pmatrix},
\eea
where the bar is used as a reminder that we are now in the $SU(2)_L\otimes SU(2)_R$ basis.~These decompose under the custodial $SU(2)_C$ as $({\bf 2,\bar{2}})={\bf 1\oplus 3}$ and $({\bf 3,\bar{3}})={\bf 1\oplus 3\oplus 5}$ providing a classification of mass eigenstates in the custodial basis after EWSB.~As a consequence of supersymmetry, each custodial representation has both a scalar and a pseudo scalar component in contrast to the GM model which has only one or the other.~Thus, after EWSB in the SCTM, we have a Higgs scalar spectrum~\cite{Cort:2013foa} which in general is significantly more complex than the spectrum found in the GM model~\cite{Gunion:1989ci}.~Of course there are also the superpartner Higgsino fermions, but these will be discussed in more detail below. 

The manifestly $SU(2)_R\times SU(2)_L$ symmetric superpotential can be written in terms of $\bar{H}$ and $\bar\Delta$ as\,\footnote{The anti-symmetric dot product is defined as $X\cdot Y = \epsilon^{ab} \epsilon_{ij} X_a^i Y_b^j $ where $\epsilon^{12} = - \epsilon_{12} = 1$ and the lower indices are acted on by $SU(2)_R$ while the upper ones are acted on by $SU(2)_L$.
},
\bea\label{eq:wo}
W_0 &=& \lambda \bar{H}\cdot \bar{\Delta}  \bar{H} +
\frac{\lambda_\Delta}{3} \Tr{\Deltab\Deltab\Deltab}  \\
&+& \frac{\mu}{2} \bar{H}\cdot \bar{H} +
\frac{\mu_\Delta}{2} \Tr{\Deltab\Deltab}, \nonumber
\eea
where $\lambda$ and $\lambda_\Delta$ are dimensionless while $\mu$ and $\mu_\Delta$ have dimensions of mass.~This gives the F-term potential,
\bea
\label{eq:vsctm}
V_{\rm{F}}&=& 
 \mu^2\,\Hb^\dagger\Hb + \mu_{\Delta}^2 \Tr{\Deltab^\dagger\Deltab}  
+ 2\lambda\mu\left(\Hb^\dagger\Deltab\Hb + c.c \right) \nn
&+& 
\lambda^2\left( 4\Tr{(\Deltab\Hb)^\dagger\,\Deltab\Hb}+(\Hb^\dagger\Hb)^2-\frac{1}{4}\,|\Hb\cdot\Hb|^2\right) \nn
&+& \lambda_\Delta^2 \left(\, \Tr{\Deltab^\dagger\Deltab^\dagger\Deltab\Deltab}
-\frac{1}{4}\,\Tr{\Deltab^\dagger\Deltab^\dagger}\Tr{\Deltab\Deltab} \right) \\
&+& \lambda\lambda_\Delta\left(\Hb\cdot\Deltab^\dagger\Deltab^\dagger\Hb
-\frac{1}{4}\Hb\cdot\Hb\,\Tr{\Deltab^\dagger\Deltab^\dagger} + c.c. \right) \nn
&+& \lambda\mu_\Delta \,\left( \Hb\cdot\Deltab^\dagger\Hb + c.c.\right) + \lambda_\Delta\mu_\Delta\,\left(\Tr{\Deltab^\dagger\Deltab^\dagger\Deltab} + c.c. \right) .\nonumber
\eea
The soft supersymmetry breaking terms are also constructed to respect the global $SU(2)_R\times SU(2)_L$,
\bea\label{eq:vsoft}
V_{\mathrm{soft}} &=& 
m_H^2 {\bar{H}}^\dagger {\bar{H}} + m_\Delta^2 \Tr{\Deltab^\dagger\Delta} \nonumber\\
&+& \Big( \frac{1}{2} B \bar{H}\cdot \bar{H} + \frac{1}{2}B_\Delta \Tr{\Deltab\Deltab} \\
&+& A_\lambda \bar{H}\cdot \bar{\Delta} \bar{H}
+  \frac{1}{3}A_{\Delta} \Tr{\Deltab\Deltab\Deltab} + h.c.\Big) ,\nonumber
\eea
where all parameters have dimensions of mass, except $B$ and $B_\Delta$ which have dimension mass squared and can be positive \emph{or} negative.~As we discuss more below, it is $B$ and $B_\Delta$ which lead to decoupling of the \emph{non}-GM spectrum in the limit they are taken large.

Along with the D-terms\,\footnote{The D-terms~\cite{Cort:2013foa} break the global $SU(2)_R\times SU(2)_L$, but as we will see below, this breaking is suppressed by the hypercharge coupling squared and only enters into \emph{non}-GM scalar masses which will be decoupled.~Custodial violating operators are introduced when the non-GM spectrum is integrated out, but these small effects are neglected in our tree level study.~A more general loop level study of the SGM is left to ongoing work~\cite{followup}.
},~\eref{vsctm} and~\eref{vsoft} give a scalar potential which leads to a mass spectrum that can be (approximately) classified~\cite{Cort:2013foa} in representations of the custodial $SU(2)_C$ after EWSB in a similar manner to the GM model~\cite{Hartling:2014zca}.~This SCTM potential leads to a scalar spectrum that contains the same scalars as found in the GM model.~In particular after rotating to the mass basis we have a $CP$ even custodial fiveplet, $H_5$,~a $CP$ odd triplet, $H_3$ (orthogonal to the Goldstones), and three Goldstone bosons which are eaten by the $W$ and $Z$ vector bosons.~There are also two real singlets, $(H_1, H_1^{'})$ which in general mix leading to the SM like Higgs $h$ and a heavy CP even scalar $H$ after rotating to the mass basis.~However, since the spectrum has now been complexifed by supersymmetry, there is now for every scalar in the GM model an additional pseudo scalar and for every pseudo scalar a new scalar.~We discuss this non-GM Higgs sector and how to decouple it in more detail below.

Using the minimization conditions~\cite{Cort:2013foa} after EWSB allows us to eliminate two parameters in the scalar potential, which we take to be the soft masses $m_H^2$ and $m_\Delta^2$ in~\eref{vsoft}, in favor of one electroweak doublet VEV ($v_H$) and one triplet VEV ($v_\Delta$) which are defined by,
\bea\label{eq:vevdef}
\langle H_1^0 \rangle &=& \langle H_2^0 \rangle  = \frac{v_H}{\sqrt{2}},\nn
\langle \chi^0 \rangle &=& \langle \phi^0 \rangle
= \langle \psi^0 \rangle  = \frac{v_\Delta}{\sqrt{2}},\\
v^2 &=& 2v_H^2 + 8 v_\Delta^2 = \frac{4 m_W^2}{g^2} . \nonumber
\eea
After rotating from the electroweak basis to the custodial mass basis~\cite{Cort:2013foa}, this gives for the masses of the GM-like scalars $H, H_3, H_5$ as well as the SM-like Higgs boson $h$,
\bea
\label{eq:MSGM}
m^{2}_{5} &=&
\frac{v_H^2 \big[\lambda(2\mu - \mu_{\Delta}) - A_\lambda \big]}{\sqrt{2}v_\Delta}
+ \frac{3}{2} \lambda v_H^2(\lambda_\Delta - 2\lambda ) \nn
&+& \sqrt{2}v_\Delta(3\lambda_\Delta\mu_\Delta+A_\Delta)-v_\Delta^2\lambda_\Delta^2 , \nn
m_{3}^2 &=&
\frac{v_H^2 \big[\lambda(2\mu - \mu_{\Delta}) - A_\lambda \big]}{\sqrt{2}v_\Delta}
+ \frac{\lambda}{2}(v_H^2 + 4 v_\Delta^2)(\lambda_\Delta - 2\lambda ) \nn
&+& 2\sqrt{2} v_\Delta \big[\lambda(2\mu - \mu_{\Delta}) - A_\lambda \big] , \\
m_{11}^2 &=& 3\lambda^2 v_H^2 , \nn
m_{12}^2 &=&  \sqrt{3}v_H\big[\lambda(\mu_\Delta -2\mu ) + A_\lambda
 +\sqrt{2} \lambda v_\Delta(3\lambda - \lambda_\Delta)\big] , \nn
m_{22}^2 &=&  \frac{v_H^2\big[\lambda(2\mu - \mu_\Delta) - A_\lambda \big]}{\sqrt{2}v_\Delta} \nn
&-& \frac{v_\Delta(3\lambda_\Delta\mu_\Delta+A_\Delta)}{\sqrt{2}} + 2v_\Delta^2\lambda_\Delta^2 ,
  \nonumber
\eea
where $m_{11},\,m_{12},\,m_{22}$ are the entries of the $2\times2$ custodial singlet mass matrix which must be diagonalized~\cite{Cort:2013foa} and where $\mathcal{M}_{12}^2 = \mathcal{M}_{21}^2$.~One of the two eigenvalues is identified with the mass of a SM-like Higgs $m^2_h$, while the other is identified with a second neutral $CP$ even scalar with mass $m^2_H$ and can be greater or less than $m_h^2$.

To summarize, we have 10 free parameters in the Higgs potential of the SCTM shown in~\eref{vsctm} and~\eref{vsoft}.~These are given by 2 quartic and 2 mass terms coming from the superpotential $(\lambda,\,\lambda_\Delta,\,\mu,\,\mu_\Delta)$ and 6 soft supersymmetry breaking mass parameters $(m_H^2,\,m_\Delta^2,\,B,\,B_\Delta,\,A_\lambda,\,A_\Delta)$ where we can also use the minimization conditions to eliminate two of the parameters in favor $v_H$ and $v_\Delta$, the doublet and triplet VEVs.

\subsection{Mapping to Georgi-Machacek (GM) Model}\label{sec:GMmodel}

In the GM model~\cite{Georgi:1985nv,Chanowitz:1985ug,Gunion:1989ci,Gunion:1990dt}, only two $SU(2)_L$ triplet scalars are added one of which is complex with hypercharge $Y=1$ (along with its conjugate) and another with hypercharge $Y=0$ which is now real in contrast to the SCTM.~Thus there are now half the number of degrees of freedom as compared to the SCTM Higgs sector.~Again the electroweak doublet and triplet fields can be arranged into the bi-doublet $({\bf 2,\bar2})$ and bi-triplet $({\bf 3,\bar3})$ representations of the global $SU(2)_L\otimes SU(2)_R$.

In fact we can easily obtain the GM model Higgs fields starting from the electroweak triplet and doublet SCTM fields in~\eref{H1H2} and~\eref{Sigmas} (see Appendix for more commonly used conventions) and imposing the conditions\,\footnote{
This then leads to the substitutions for operators in the scalar potential:
$\Hb\cdot\Hb \rightarrow - \Hb^\dagger\Hb,\,~\Hb\cdot\Deltab^\dagger\Hb \rightarrow - \Hb^\dagger\Deltab\Hb$.
}, 
\bea
\label{eq:DelHmap}
\bar\Delta^\dagger = \bar\Delta,~H_2 = -i \sigma_2 H_1^\ast \, .
\eea
Furthermore, with these conditions we not only recover the GM model Higgs fields~\cite{Hartling:2014zca}, but we can also derive the Higgs potential of the GM model from the SCTM Higgs potential, $V_{\rm{SCTM}} 
\equiv V_{\rm{F}} + V_{\rm{soft}}$ (\eref{vsctm} plus~\eref{vsoft}).

To see this we first apply the constraints\,\footnote{The contributions from the D-terms~\cite{Cort:2013foa} to the potential all vanish when we impose the constraints in~\eref{DelHmap} and thus only enter into \emph{non}-GM scalar masses to be discussed more below.} in~\eref{DelHmap} to $V_{\rm{SCTM}}$, after which the Higgs potential is written as,
\bea\label{eq:vgm}
V_{\rm{GM}} &=& 
\frac{1}{2}\mu_2^2\,{\Hb}^\dagger {\Hb} + 
\frac{1}{2}\mu_3^2\, \Tr{\Deltab\Deltab} + \lambda_1 ({\Hb}^\dagger {\Hb})^2  \nn
&+&  (\lambda_2 + \frac{1}{4}\lambda_5) ({\Hb}^\dagger {\Hb}) 
\Tr{\Deltab\Deltab} - 2\,\lambda_3 \Tr{({{\bar{\Delta}}}{{\bar{\Delta}}})^2 }\nn
&+&\, (\frac{3}{2}  \lambda_3 + \lambda_4) \,\Tr{\Deltab\Deltab}^2 - \lambda_5\, \Hb^\dagger\Deltab\Deltab \Hb \\
&+&  \frac{M_1}{2}\Hb^\dagger\Deltab\Hb +2M_2\,\Tr{\Deltab\Deltab\Deltab} .\nonumber
\eea
When expressed in terms of the component electroweak fields in~\eref{H1H2} and~\eref{Sigmas} (with the condition in~\eref{DelHmap} enforced),~\eref{vgm} matches precisely the GM model Higgs potential\,\footnote{This includes the $\mathcal{Z}_2$ breaking mass parameters $M_1$ and $M_2$, thus allowing for a proper decoupling limit to exist~\cite{Hartling:2014zca}.}
given in~\cite{Hartling:2014zca} (and Appendix).

Comparing coefficients of each operator in $V_{\rm{GM}}$ with those in $V_{\rm{SCTM}}$ \emph{after}~\eref{DelHmap} has been applied allows us to obtain a mapping between the Higgs potential parameters of the two models,
\bea
\label{eq:relation}
\lambda_1 &=& \frac{3}{4} \lambda^2,~ 
\lambda_2  = \lambda^2,~ 
\lambda_3 = -\frac{1}{2} \lambda_\Delta^2,\nn
\lambda_4 &=& \frac{1}{2} \lambda_\Delta^2,~
\lambda_5 = 2\lambda (\lambda_\Delta -2\lambda ),\nn
M_{1} &=& 4\big[\lambda(2\mu - \mu_{\Delta}) - A_\lambda \big],\\
M_{2} &=& \frac{1}{3}(3\lambda_\Delta\mu_\Delta + A_\Delta ), \nn
\mu_2^2 &=& 2(\mu^2 + m_H^2) + B, \nn
\mu_3^2 &=& 2(\mu_\Delta^2 + m_\Delta^2) + B_\Delta. \nonumber
\eea
We can use this mapping to define the SGM model Higgs potential in terms of~\eref{vgm} with~\eref{relation} imposed or, equivalently, as $V_{\rm{SCTM}}$ with the constraint in~\eref{DelHmap} applied.~One can also verify that imposing the relations in~\eref{relation} on the GM scalar masses in~\cite{Hartling:2014zca} reproduces exactly the GM-like scalar masses in~\eref{MSGM} once, using the vacuum conditions~\cite{Cort:2013foa}, $m_H^2$ and $m_\Delta^2$ have been eliminated in favor of the VEVs $v_H$ and $v_\Delta$.

The mapping between the SGM and GM model Higgs potential parameters in~\eref{relation} also implies the following constraints \emph{between} the five dimensionless quartic couplings in the GM model Higgs potential,
\bea
\label{eq:gmcon}
\lambda_1 &=& \frac{3}{4} \lambda_2,\,\lambda_3 = -\lambda_4,\\
\lambda_5 &=& -4\lambda_2 + 2\sqrt{2\lambda_2\lambda_4} . \nonumber
\eea
Thus we see the five quartic couplings in the GM model Higgs potential can be written in terms of only $\lambda_2$ and $\lambda_4$.~This defines a `constrained' GM model in terms of the GM Higgs potential in~\eref{vgm} with~\eref{gmcon} imposed.~These constraints imply correlations between operators in the Higgs potential of the GM model and could be a signal of its supersymmetric origin.~Note that the conditions in~\eref{gmcon} satisfy the constraints for the GM potential being bounded from below~\cite{Hartling:2014zca} as expected for a theory with a supersymmetric origin.~Note also that~\eref{relation} and holomorphy of the superpotential implies the bound on the quartic couplings $0 < \lambda_{2,4} \in \Re$.

These constraints could manifest in correlations between rate measurements as well as perhaps in differential distributions for precisely measured channels such as the $4\ell$ and $2\ell\gamma$ final states.~These may serve as useful additional probes, over a range of center of mass energies~\cite{Chen:2014ona,Chen:2015rha,Chen:2016ofc,Stolarski:2016dpa,Vega-Morales:2017pmm}, for distinguishing between the SGM and GM models at the LHC, particularly once large data sets are collected at a high luminosity LHC.~As we'll see below, the GM limit of the SCTM also implies large $\mu_{2,3}^2$ mass parameters in the GM model.~So we see the sign of a supersymmetric origin for a GM-like model could be correlated quartic couplings along with large $\mu_{2,3}^2$.

One could in principle tune the GM Higgs potential parameters to satisfy these constraints.~This would then lead to a `slice' in parameter space where the GM model can generically give very similar signals to the SGM model.~To ascertain a true `smoking gun' signal of its supersymmetric origin will require observing effects from the superparter sector and in particular the (light) fermionic sector.~These include both tree level and one loop effects which we examine in more detail below.

\subsection{The `GM limit' of the SCTM}\label{subsec:GMlimit}

The non-GM scalars in the SCTM~\cite{Cort:2013foa}, which we will refer to as the `\emph{mirror}-GM' Higgs sector, are comprised of two (neutral) $CP$-odd custodial singlet psuedo scalars, two $CP$ even triplet scalars, and a $CP$-odd fiveplet pseudo scalar.~The two triplet scalars can in general mix as can the two pseudo scalar singlets.~The fiveplet pseudo scalar, like its $CP$ even counterpart (as found in the GM model), is prevented by custodial symmetry from mixing at tree level.~However, in contrast to the $CP$ even fiveplet it does \emph{not} have tree level couplings\,\footnote{Note the physical $T$-odd scalar in Littlest Higgs Models with $T$-parity~\cite{Cheng:2003ju,Cheng:2004yc,Low:2004xc,Hubisz:2004ft} resembles the custodial fiveplet pseudo scalar.} to $WW, ZZ$, or $WZ$ pairs.~Note that the neutral and charged MSSM like Higgs scalars are contained within the custodial triplet scalars and singlet pseudo scalars, which as we'll see are decoupled in the GM limit.

To see how we can decouple the \emph{mirror}-GM scalars \emph{without} decoupling the GM-like scalars in~\eref{MSGM}, we examine their masses~\cite{Cort:2013foa} after expanding around $v_\Delta \approx 0$,
\bea
\label{eq:MMGMlim}
M^{2}_{5} &\approx&
\frac{ v_H^2 [\lambda(2\mu - \mu_{\Delta}) - A_\lambda]}{\sqrt{2} v_\Delta} - 2 B_{\Delta}  \nn
&+& \frac{1}{2} \lambda  v_H^2(\lambda_\Delta - 6\lambda) + \mathcal{O}(v_\Delta) , \nn
M_{3^\prime}^2 &\approx&
\frac{ v_H^2 [\lambda(2\mu - \mu_{\Delta}) - A_\lambda]}{\sqrt{2} v_\Delta} - 2 B_{\Delta} \nn
&+& \frac{1}{2} \lambda  v_H^2(3\lambda_\Delta - 2\lambda) + \mathcal{O}(v_\Delta),\\
M_{3}^2 &\approx&
\frac{1}{2} v_H^2 (G^2 + 2\lambda^2) +  2 B ,\nn
M_{1^\prime}^2 &\approx& 
\frac{v_H^2 [\lambda(2\mu - \mu_{\Delta}) - A_\lambda]}{\sqrt{2}v_\Delta} - 2 B_{\Delta} \nn
&+& 2 \lambda \lambda_\Delta v_H^2 + \mathcal{O}(v_\Delta) , \nn
M_{1}^2 &\approx& 2 B , \nonumber
\eea
where we use similar notation to~\eref{MSGM} in order to denote custodial singlet, triplet, or fiveplet, but use $M$ insead of $m$.~For the custodial triplet mass we have $G^2 = g^2 + g^{\prime2}$ for the neutral component and $G^2 = g^2$ for the charged\,\footnote{Since $G^2$ differs between the neutral and charged components, custodial symmetry is broken at tree level.~However, these effects are suppressed by the small hypercharge coupling squared~\cite{Cort:2013foa} and only affect the SGM model Higgs spectrum at loop level.}.~The small $v_\Delta$ limit is not necessary for our analysis, but simplifies the discussion below.

What is crucial to note is that unlike for the GM-like scalar masses in~\eref{MSGM}, the masses in~\eref{MMGMlim} depend explicitly (linearly) on the soft supersymmetry breaking masses $B$ and $B_\Delta$.~This opens the possibility of decoupling \emph{all} non-GM scalars while ensuring the GM like scalars remain light and around the weak scale.~In particular, if we take $B > 0, B_\Delta < 0$, while holding $v_\Delta$ and the other scalar potential parameters fixed, then in the limit $|B|,|B_\Delta | \to \infty$, all \emph{mirror}-GM scalar masses in~\eref{MMGMlim} become large while the GM-like scalars in~\eref{MSGM} are unaffected and remain light.~As in the MSSM~\cite{Martin:1997ns}, taking $B\to\infty$ decouples the (custodial) MSSM like scalars with masses $M_1$ and $M_3$ in~\eref{MMGMlim}.~Conversely, one can obtain the MSSM by taking the $|B_\Delta|, |\mu_\Delta| \to \infty$ limit.

There are subtleties in ensuring the decoupling behavior needed to obtain the GM Higgs sector at low energies.~To examine this, we would like to find, purely in terms of Higgs potential parameters (instead of VEVs), the limit where all  \emph{mirror}-GM Higgs bosons decouple while all GM-like Higgs bosons, as well as the SM-like Higgs boson, are left light and around the weak scale.~In taking the large $|B_\Delta |$ limit for the $M_1^\prime, M_3^\prime, M_5$ masses in~\eref{MMGMlim} we have assumed implicitly that $v_\Delta$ does not go to zero as $|B_\Delta |$ is taken large, which would cause the GM-like scalars to decouple as well.

To gain insight for how this is possible we first note that the vacuum conditions~\cite{Cort:2013foa} (in the small $v_\Delta$ limit) impose the constraint on the Higgs potential parameters,
\bea
\label{eq:vaclim}
\frac{v_H^2 \left(\lambda \left(2 \mu - 
\mu _{\Delta } \right) - A_\lambda \right)}{\sqrt{2} v_{\Delta }}
\approx
B_{\Delta } + m_{\Delta }^2  
+ \mu _{\Delta }^2 \\
+~\lambda v_H^2 \left(3 \lambda -\lambda _\Delta \right) 
+ \mathcal{O}(v_\Delta)\,,\nonumber
\eea
where we see the ratio on the left appears explicitly in the masses of custodial scalars which originate from electroweak triplets ($m_H,\,m_3,\,m_5,\,M_1^\prime,\,M_3^\prime,\,M_5$).~Substituting~\eref{vaclim} into the masses in~\eref{MSGM} and~\eref{MMGMlim} gives for the GM-like scalar masses (for small $v_\Delta$),
\bea
\label{eq:MSGMlim2}
m^{2}_{5} &\approx&
 B_{\Delta} + m^2_{\Delta} + \mu^2_{\Delta} 
+ \frac{1}{2} \lambda \lambda_\Delta v_H^2 , \nn
m_{3}^2 &\approx&
B_{\Delta} + m^2_{\Delta} + \mu^2_{\Delta} 
+ \frac{1}{2} \lambda (4\lambda - \lambda_\Delta) v_H^2 , \\
m_{H}^2 &\approx& 
B_{\Delta} + m^2_{\Delta} + \mu^2_{\Delta} 
+ \lambda (3\lambda - \lambda_\Delta) v_H^2 , \nn
m_{h}^2 &\approx&  3\lambda^2 v_H^2 , \nonumber
\eea
where now the doublet (and triplet) VEVs are \emph{dependent} parameters and fixed once the Higgs potential parameters are fixed.~For the \emph{mirror}-GM scalars this gives,
\bea
\label{eq:MMGMlim2}
M^{2}_{5} &\approx&
 - B_{\Delta} + m^2_{\Delta} + \mu^2_{\Delta} 
- \frac{1}{2} \lambda \lambda_\Delta v_H^2 , \nn
M_{3^\prime}^2 &\approx&
- B_{\Delta} + m^2_{\Delta} + \mu^2_{\Delta} 
+ \frac{1}{2} \lambda (4\lambda + \lambda_\Delta) v_H^2 , \nn
M_{3}^2 &\approx&
\frac{1}{2} v_H^2 (G^2 + 2\lambda^2) +  2 B,\\
M_{1^\prime}^2 &\approx& 
- B_{\Delta} + m^2_{\Delta} + \mu^2_{\Delta} 
+ \lambda (3\lambda + \lambda_\Delta) v_H^2 , \nn
M_{1}^2 &\approx& 2 B. \nonumber
\eea
Examining~\eref{MSGMlim2} and~\eref{MMGMlim2} we can see what is the necessary limit of lagrangian parameters to decouple the \emph{mirror}-GM sector while keeping the GM-like scalars as well as the (mostly) SM Higgs boson light.~In particular, by imposing the conditions $B_\Delta \approx - (m_\Delta^2 + \mu_\Delta^2) < 0$, we can ensure the GM-like masses do \emph{not} decouple along with the \emph{mirror}-GM masses in the $|B_\Delta| \to \infty$ limit.

We expect the two soft supersymmetry breaking parameters to be of the same order so $|B_{\Delta }| \sim |m_{\Delta }^2|$ while the Higgs triplet superpotential mass parameter $\mu_\Delta$ is \emph{a priori} unrelated to the soft breaking parameters.~We take it to be at around the same scale as the doublet mass parameter $\mu$, but below $B, B_\Delta$.~Thus, we find the limit for decoupling \emph{only} the \emph{mirror}-GM sector to be $B_\Delta \approx - \, m_\Delta^2 < 0$ and taking $m_\Delta^2, B \to \infty$ with all other mass scales fixed.~This gives for the SGM scalar masses,
\bea
\label{eq:MGMlim3}
m^{2}_{5} &\approx&
\mu^2_{\Delta} + \frac{1}{2} \lambda \lambda_\Delta v_H^2 , \nn
m_{3}^2 &\approx&
\mu^2_{\Delta} + \frac{1}{2} \lambda (4\lambda - \lambda_\Delta) v_H^2 , \\
m_{H}^2 &\approx& 
\mu^2_{\Delta} + \lambda (3\lambda - \lambda_\Delta) v_H^2 , \nn
m_{h}^2 &\approx&  3\lambda^2 v_H^2 , \nonumber
\eea
while all of the \emph{mirror}-GM scalars become very heavy and decouple.~We see also that with these conditions, the ratio on the left hand side in~\eref{vaclim} remains finite.~This implies $v_\Delta$ can be held fixed and does not necessarily go to zero as $|B_\Delta| \to \infty$.~We thus see it is indeed possible to decouple the \emph{mirror}-GM scalars while ensuring the GM-like scalars remain light at around the weak scale.~Of course formally we cannot take $m_\Delta^2, B \to \infty$ without reintroducing the quadratic divergences inherent in the GM model~\cite{Gunion:1990dt}, and thus a fine tuning, so in practice we only require $m_\Delta^2, B \gg v$ where $v$ is the electroweak scale.~As we will see below, already for $m_\Delta^2, B \approx 1$~TeV one obtains the GM scalar spectrum at the weak scale to a very good approximation and without severe tuning.~A more precise analysis requires a loop level analysis and would be interesting to pursue, but is beyond the scope of the current study and left to future work.

We see also in~\eref{MGMlim3} that the limit $\mu^2_{\Delta} \to \infty $ decouples the GM-like scalars and maps onto the $M_{1},\,M_{2},\,\mu^2_3  \to \infty $ decoupling limit in the GM model~\cite{Hartling:2014zca}.~Furthermore, using~\eref{relation} we see that the GM limit $B_\Delta, B \to \infty$ leads to $\mu_{2,3}^2 \to \infty$ in the GM model.~This does not necessarily decouple the GM scalar sector however, as these effects can be compensated for with large $M_{1,2}$ mass parameters.~We also note that the `un-complexification' constraint in~\eref{DelHmap}, which leads to the mapping between the SCTM and GM models, breaks supersymmetry.~This is of course consistent with the GM limit of the SCTM defined by the large soft masses.

Finally, we point out that if we instead take $B_\Delta \approx + \, m_\Delta^2 > 0$, we can obtain the inverted spectrum by taking the limit $|B_\Delta| \to \infty$ while keeping $B$ at the weak scale (see~\eref{MSGMlim2} and~\eref{MMGMlim2}).~In this case now the GM-like scalars are heavy while the \emph{mirror}-GM scalars, plus the SM Higgs boson, are light and around the weak scale.~This limit, which we refer to as the \emph{mirror}-GM model, has interesting phenomenology that can also mimic the GM model with certain important differences to be examined in ongoing work~\cite{followup}.

\subsection{(Custodial) fermion superpartners - LSP}\label{sec:fermions}

Of course being a supersymmetric theory, the SGM contains fermionic superpartners which can in principle also be light.~As the masses of these fermions is taken large their effects decouple and the SGM phenomenology looks very similar to GM phenomenology.~As they  become light they can effect the decays of all the GM-like scalars in the SGM at tree-level and one loop.

There are three contributions to the fermion custodial superparter sector, the first two of which are as in the scalar case, from the electroweak Higgs doublet and triplet chiral superfields.~The five neutral Higgsinos and two gauginos lead to a $7\times 7$ neutralino mass matrix in the SCTM which generically is non-trivial~\cite{Cort:2013foa}.~However, since the fields in~\eref{H1H2} and~\eref{Sigmas} are chiral superfields, the same rotations which take the Higgs scalars~\cite{Cort:2013foa} into the custodial basis, also rotate the fermion fields and greatly simplify the mass matrices.~These lead to two Higgsino custodial singlets, two triplets, and one fiveplet which constitute the superpartners of the SGM and \emph{mirror}-GM custodial scalars.~The third contribution to the fermion superpartners comes from the hypercharge and $SU(2)_L$ gauge vector superfields.~If we neglect hypercharge interactions \emph{or} assume universal gaugino masses, the electroweak gauginos decompose, like the electroweak gauge bosons, into custodial singlet and triplet representations.~Therefore, they also mix at tree level with Higgsinos in the same custodial representation.~In the end we are left with three (approximately) custodial singlets $(\tilde{h}_1,\tilde{\delta}_1,\tilde{\gamma})$, three triplets $(\tilde{Z}, \tilde{h}_3^0, \tilde{\delta}_3^0)$, and one fiveplet ($\tilde{\delta}^0_5$).~The lightest stable particle (LSP) makes a potential dark matter candidate~\cite{Delgado:2015aha} and is formed out of some combination of the neutral components of the custodial fermions.~In this basis,
\bea
\label{eq:Psi0_vector}
\Psi^0 &=& \Big(\tilde{h}_1, \tilde{\delta}_1, \tilde{\gamma}, \tilde{Z}, \tilde{h}_3^0, \tilde{\delta}_3^0, \tilde{\delta}^0_5 \Big) ,
\eea
the neutralino mass matrix simplifies to the (almost) block diagonal form given by,
\begin{widetext}
\small
\bea
\label{eq:MF0}
M^0_F =
	\begin{pmatrix}
	 \frac{3}{\sqrt{2}} \lambda v_\Delta - \mu & \sqrt{3} \lambda v_H & 0 & 0 & 0 & 0 & 0 \\
	 \sqrt{3} \lambda v_H & -\sqrt{2}\lambda_\Delta v_\Delta + \mu_\Delta & 0 & 0 & 0 & 0 & 0 \\
	 0 & 0 &  \frac{g^2 M_{\tilde{B}} + {g^\prime}^2 M_{\tilde{W}} }{g^2 + {g^\prime}^2} & \frac{g{g^\prime} (M_{\tilde{W}} - M_{\tilde{B}})}{g^2 + {g^\prime}^2}  & 0 & 0 & 0 \\
	 0 & 0 &  \frac{g{g^\prime} (M_{\tilde{W}} - M_{\tilde{B}})}{g^2 + {g^\prime}^2}  &  \frac{{g^\prime}^2 M_{\tilde{B}} + g^2 M_{\tilde{W}}}{g^2 + {g^\prime}^2}  & \sqrt{\frac{1}{2}(g^2 + {g^\prime}^2)}\,v_H & \sqrt{2(g^2 + {g^\prime}^2)}\,v_\Delta & 0 \\
	 0 & 0 & 0 & \sqrt{\frac{1}{2}(g^2 + {g^\prime}^2)}\,v_H & \frac{1}{\sqrt{2}} \lambda v_\Delta + \mu & -\sqrt{2}\lambda v_H & 0 \\
	 0 & 0 & 0 & \sqrt{2(g^2 + {g^\prime}^2)}\,v_\Delta & -\sqrt{2}\lambda v_H & \frac{1}{\sqrt{2}} \lambda_\Delta v_\Delta - \mu_\Delta & 0 \\
	 0 & 0 & 0 & 0 & 0 &0 &  \frac{1}{\sqrt{2}} \lambda_\Delta v_\Delta + \mu_\Delta \\
	 
	\end{pmatrix} ,~~~~~~
\eea
\normalsize  
\end{widetext}
where $M_{\tilde{B}}$ and $M_{\tilde{W}}$ are the supersymmetry breaking electroweak bino and wino masses respectively~\cite{Cort:2013foa}.

We see that we are only prevented from having the mass matrix in block diagonal form by the zino-photino ($\tilde Z - \tilde\gamma$) mixing represented in the 34 and 43 matrix entries.~Note also that, if we want to keep the $\mu$ terms around the weak scale, only the photino and zino can be decoupled by taking the gaugino masses large while all other contributions go to zero in the absence of EWSB.~Since they mix, decoupling the zino also decouples part of the custodial triplet Higgsinos.~The rest of the custodial fermions are decoupled as $\mu,\,\mu_\Delta\to \infty$.

In the limit we take the hypercharge coupling $g^\prime\to 0$, in which case custodial symmetry (neglecting quark and lepton Yukawa sectors) is exact at tree level, this mixing goes to zero and the neutralino mass matrix becomes exactly block diagonal.~We can then decompose the neutralino mass matrix into a $2\times2$ sub matrix for the Higgsino custodial singlets ($\tilde{h}_1, \tilde{\delta}_1$), a $1\times1$ for the custodial singlet photino ($\tilde\gamma$), a $3\times3$ for the three custodial triplets ($\tilde{Z}, \tilde{h}_3^0, \tilde{\delta}_3^0$) which in general mix, and finally a  $1\times1$ for the fiveplet ($\tilde{\delta}^0_5$) higgsino.~In particular, in this limit the sub mass matrices for the neutral components of the custodial triplets will be equal to their corresponding charged ones as required by custodial symmetry.~Note we could have also put the mass matrix in block diagonal form by taking universal gaugino masses $M_{\tilde{B}} = M_{\tilde{W}}$.~However, in this case one still has custodial breaking effects due to the hypercharge couplings $g^\prime$ entering the custodial triplets which manifests as an $\mathcal{O}({g^\prime}^2)$ splitting between the masses of the neutral and charged components.

At tree level the custodial fiveplet Higgsino has degenerate neutral, singly charged, and doubly charged components.~Small custodial breaking hypercharge interactions enter into 1-loop corrections of the custodial fiveplet mass and break the degeneracy leading to the lightest component being the neutral one.~This is crucial for ensuring the charged components can decay and avoid problems with the many stringent experimental constraints on charged stable particles.~This is similarly true about the custodial triplet fermions, but in addition, these have a tree level splitting due to the small breaking of custodial symmetry by hypercharge interactions entering through the $D$-terms~\cite{Cort:2013foa}.~These introduce the $\mathcal{O}({g^\prime}^2)$ corrections into the neutral component masses seen in~\eref{MF0}.~Furthermore, over large regions of parameter space, the lightest neutral component of these fermions can make a viable thermal dark matter candidate~\cite{Delgado:2015aha}.

\section{GM versus SGM model at LHC}\label{sec:SGM}

In this section we compare phenomenology between the SGM model and the `constrained' GM model, as defined by~\eref{vgm} with~\eref{gmcon} imposed, and identify signals which might be useful for distinguishing them.~For our analysis we go back to the scalar masses of~\eref{MSGM} where we have eliminated $m_\Delta^2$ and $m_H^2$ in favor of the doublet and triplet VEVs, $v_H$ and $v_\Delta$, so they are again independent parameters.~On the GM side we implicitly use the vacuum conditions~\cite{Hartling:2014zca} to eliminate $\mu_2^2$ and $\mu_3^2$ again in favor of the doublet and triplet VEVs, $v_\phi$ and $v_\chi$ respectively, which map onto the SGM VEVs, $v_H$ and $v_\Delta$.

For the remainder of this study, we define the SGM model as the GM limit of the SCTM given by,
\bea\label{eq:gmlim}
B = - B_\Delta \gg M^2 \,~\Rrightarrow \rm{SCTM \to SGM}\, ,
\eea
where $M \equiv \mu,\, \mu_\Delta, A_\lambda, A_\Delta, v_H, v_\Delta$ represents all other mass scales present in the SCTM Higgs potential.~As we'll see below, numerically it turns out that for $B \sim (\rm{TeV})^2$ and all other mass scales $M \sim v = 246$~GeV, one already begins to obtain \emph{only} the GM spectrum at the weak scale.~For our scans we set explicitly $B = (1\,\rm{TeV})^2$ as well as the gaugino masses $M_{\tilde{B}} = M_{\tilde{B}} = 1$\,TeV.

\subsection{`Smoking guns' signals of SGM versus GM}\label{sec:diboson}

There are a number of possible `smoking gun' signals of the SGM which could be used to establish the supersymmetric nature of a GM-like model.~Of course more generally when~\eref{gmcon} is not satisfied, the GM and SGM will have very different phenomenology at the LHC.~In particular, though many of the LHC signals would be similar, their magnitudes and the correlations among them would not be the same.~This may also manifest itself (via tree or 1-loop effects) in differential distributions which could perhaps be observable in the precisely measured $4\ell$ and $2\ell\gamma$ channels~\cite{Chen:2014ona,Chen:2015rha,Chen:2016ofc,Vega-Morales:2017pmm}.~As we'll see, the `slice' in parameter space of the GM model represented by the constraints in~\eref{gmcon} can closely mimic the SGM model depending on the fermion superpartner masses.

The smoking gun of GM type models is the presence of the custodial fiveplet~\cite{Logan:2017jpr,Zhang:2017och} scalar.~Perhaps the most well studied signal is the decay of the doubly charged component into same sign $W$ bosons~\cite{Georgi:1985nv,Vega:1989tt,Englert:2013wga,Chiang:2014bia}, which in turn leads to a same sign di-lepton signal, plus missing transverse energy ($E_T$) due to the final state neutrinos.~This decay is of course also present in the SGM, but now there is an additional decay through pairs of charginos which also leads to a same sign di-lepton signal plus missing $E_T$.~In this case however, the missing $E_T$ includes a pair of LSPs leading to a significantly altered missing $E_T$ spectrum.~This implies the missing $E_T$ spectrum in the doubly charged scalar decay could be used to distinguish between the GM and SGM models.~The additional decay mode through pairs of charginos may also allow for evading constraints from like-sign $W$ boson searches~\cite{Khachatryan:2014sta,deFlorian:2016spz}.~Similar considerations hold for decays of the singly charged component to $W^\pm Z$ leading to tri-lepton signals plus missing $E_T$, but we leave an exploration of these possibilities to ongoing work~\cite{followup}.

Of course even if GM-like scalars are not observed, they can affect decays of the SM-like Higgs boson at both tree level and one loop.~In both the GM and SGM models,~$h$ and $H_5$ have tree level decays to $W^{\pm}W^\mp$ and $ZZ$ vector boson pairs while, as discussed above, the charged components of $H_5$ can also decay to $W^{\pm}Z$~\cite{Zhang:2017och} and $W^{\pm}W^\pm$ pairs~\cite{Englert:2013wga}.~At one loop both $h$ and $H_5$ have decays to $Z\gamma$ and $\gamma\gamma$ while $H_5$ can also decay to $W^\pm\gamma$~\cite{Degrande:2017naf}, which leads to an interesting mono-lepton plus photon signal with missing $E_T$.~As we'll see, the primary difference between the SGM and GM decay patterns comes from the effects of the fermionic superpartners.~These can affect the total decay widths of $h$ and $H_5$ once the LSP is light enough for 2 (or three) body scalar decays.~In addition, their effects can be important at one loop and in particular for the loop in induced $W^\pm\gamma,\,Z\gamma$, and $\gamma\gamma$ decays.

With these considerations in mind, we examine in particular decays of the custodial fiveplet $H_5$ and the SM-like Higgs boson $h$ focusing on decays to the well measured $\gamma\gamma$ and $ZZ$ final states.~We examine a pair of simplified benchmark scenarios to compare between the SGM and GM models.~A more in depth study including other possible decays is left to future work as is a detailed scan to establish the allowed parameter space including direct~\cite{Logan:2015xpa,Delgado:2016arn} and indirect~\cite{Hartling:2014aga,Queiroz:2014zfa} constraints.

\subsection{Parameter space and benchmark scans}\label{sec:scans}

For our parameter scans we will also in the following assume a gauge mediated supersymmetry breaking scenario~\cite{Delgado:2015bwa} in order to fix $A = A_\Delta = 0$.~This leaves us with a dependence on only the four superpotential parameters (see~\eref{wo}) and two VEVs.~This gives a 1-to-1 mapping between the six Higgs potential parameters in the GM and SGM models, which we represent schematically as,
\bea\label{eq:parammap}
(\lambda,\,\lambda_\Delta,\,\mu,\,\mu_\Delta, v_H, v_\Delta) \Leftrightarrow
(\lambda_2,\,\lambda_4,\,M_1,\,M_2,\,v_\phi,\,v_\chi),~~~~~
\eea
where the sets of parameters correspond to $\rm{SGM} \Leftrightarrow \rm{GM}$.~The constraint $A = A_\Delta = 0$ is not strictly necessary and other supersymmetry breaking scenarios could be considered, but this assumption simplifies our current analysis without qualitatively changing the discussion.~We will utilize this mapping below to analyze and compare some of the phenomenology in the GM and SGM models for a pair of benchmark scenarios. 

We also use measurements~\cite{Agashe:2014kda} of the Higgs and $W$ boson masses as well as electroweak gauge couplings to impose the pair of additional constraints,
\bea
\label{eq:vevmh}
v
= 246\, {\rm GeV},~
m_{h} = 125 \, \mbox{GeV} .
\eea
Using~\eref{vevdef} and~\eref{MSGM}, this allows us to eliminate doublet and triplet VEVs, in both the SGM and GM models.~This reduces the six dimensional parameter space in~\eref{parammap} to the four superpotential parameters $(\lambda,\,\lambda_\Delta,\,\mu,\,\mu_\Delta)$ which in turn maps onto the four GM potential parameters $(\lambda_2,\,\lambda_4,\,M_1,\,M_2)$ using~\eref{relation}.

For the pair of benchmark scenarios considered here we impose the additional constraints,
\bea
\label{eq:points12}
\rm{{\bf point \, 1:~}}
\lambda &=& -\lambda_\Delta,~~\mu = \mu_{\Delta}, \\
\rm{{\bf point \, 2:~}}
\lambda &=& \lambda_\Delta,~~\mu = -\mu_{\Delta},\nonumber
\eea
leaving us finally wth two degrees of freedom.~For our scans we trade these in for the custodial fiveplet and triplet masses ($m_5$ and $m_3$) in terms of which all other Higgs potential parameters are then determined.~The fermion superpartner masses are then also fixed (after also fixing the gaugino masses).~After using~\eref{relation}, this then also determines the Higgs potential parameters in the GM model in terms of $m_3$ and $m_5$.

The top Yukawa coupling is fixed once $v_\Delta$ is determined by the requirement of reproducing the observed top mass~\cite{Cort:2013foa,Garcia-Pepin:2014yfa}.~Larger values of $v_\Delta$ require larger top Yukawa couplings which in turn induces sizable radiative corrections to the SM-like Higgs boson mass from stop loops.~We also note that, while it is increasingly disfavored experimentally~\cite{Chen:2016ofc}, a negative top Yukawa coupling could in principle be generated~\cite{Hedri:2013wea} in the SGM, but we do not consider this possibility here.

It has also been shown that loop corrections to the other custodial Higgs boson masses can be large, and in some cases divergent, in (non-supersymmetric) GM type models~\cite{Braathen:2017izn,Krauss:2017xpj}.~However these effects are less drastic for custodial scalars in the mass range we consider ($\lesssim 250$~GeV) so for present purposes they are neglected for simplicity since calculations of the custodial scalar masses is not the focus of this study.~However, it may be interesting, and necessary, to consider if these 1-loop corrections give additional possibilities for distinguishing between the SGM and GM models.~Loop effects from squark and sleptons are included, but their masses are set large enough ($\sim$~TeV) that these are negligible.~Lowering their masses could affect our results dramatically, but since this sector is independent of the Higgs sector we take it to be much heavier than the weak scale.~For all of the calculations needed to conduct our two dimensional scans we have used the SARAH/SPheno~\cite{Porod:2003um,Porod:2011nf,Staub:2015kfa} package and validated for a few random points with FeynArts/FormCalc/LoopTools~\cite{Hahn:1998yk,Klasen:2002xi}.

\subsection{Fiveplet decays to $\gamma\gamma$ and $ZZ$ pairs}\label{sec:h5decays}

Diphoton searches have been shown to be a powerful and robust direct search probe for (light) fermiophobic Higgs bosons~\cite{Mrenna:2000qh,Delgado:2016arn,Brooijmans:2016vro}, of which the custodial fiveplet found in GM-type Higgs models is a particular example.~This is particularly true when combined with Drell-Yan Higgs pair production~\cite{Akeroyd:1995hg,Akeroyd:2003xi,Delgado:2016arn} which can be sizable for Higgs scalars below $\sim 250$~GeV,~especially at the LHC~\cite{Akeroyd:2003bt,Delgado:2016arn}.~Four photon searches have also been shown to be useful~\cite{Akeroyd:2005pr,Aaltonen:2016fnw} in the case where the pair of Higgs bosons produced are not degenerate.

At the light masses we consider, the production of a custodial fiveplet is dominated by the Drell-Yann $pp\to H_5^0 H_5^{\pm}$ Higgs pair channel~\cite{Delgado:2016arn} which depends only on the $SU(2)_L$ gauge coupling and representations.~We have used the Madgraph~\cite{Alwall:2014hca} framework developed for the GM model in~\cite{Hartling:2014xma} to compute the pair production cross section at leading order for a 13~TeV LHC.~We have neglected single production channels which are suppressed by small VEV and/or Higgs mixing angles and contribute only $\lesssim\mathcal{O}(10\%)$ to the total production cross section once constraints from coupling measurements of the 125~GeV SM-like Higgs boson~\cite{Khachatryan:2014kca} are taken into account limiting us to $v_\Delta \lesssim 15$~GeV in our parameter scans.~Thus to a good approximation the fiveplet production is independent of $v_\Delta$ and the same in the GM and SGM models, ranging from $\mathcal{O}(0.1 - 10) \, pb$ in the mass range $50 - 250$~GeV~\cite{Delgado:2016arn}.~Current sensitivity of LHC diphoton searches~\cite{CMS-PAS-HIG-17-013} in this mass range is $\mathcal{O}(0.1)\,pb$ making diphoton searches via pair production an interesting probe of these Higgs sectors.~A more in depth investigation of light diphoton signals utilizing the Higgs pair production mechanism is ongoing~\cite{followup}.

Assuming the fiveplet is the lightest non-SM Higgs scalar, we show in~\fref{H5aaconts} contours of the production cross section times branching ratio into two photons as a function of $m_5$ versus $m_3$ in both the GM (blue solid lines with background color) and SGM models (red solid line).~While it is the light charginos which give the loop contributions to the diphoton width, we show in the black dashed lines various contours of the neutralino LSP mass in the SGM.~As discussed above, which neutralino makes up the dominant component of the LSP depends on the parameter point chosen.~Whether we have a light or heavy LSP (and by extension other fermionic superpartners) largely determines the qualitative behavior seen in the diphoton branching ratios.

This can be seen in~\fref{H5aaconts} where for benchmark point 1 (top), which leads to larger LSP masses, the GM and SGM contours are approximately alined.~This is particularly true as we go to larger LSP masses where $H_5$ can no longer decay to the LSP and its 1-loop effects are negligible.~The larger LSP masses for benchmark point 1, in which we set $\lambda = -\lambda_\Delta$, are due to cancelations between different terms in the custodial fiveplet mass (see~\eref{MGMlim3}) which requires larger $\mu_\Delta$ terms to satisfy the constraint from the input $m_5$ masses.~Since, apart from the gauginos, the fermion superpartner masses are largely determined by the $\mu$ terms when they are large, this leads to heavier LSP masses.~In this heavier LSP scenario, where we find the LSP is composed of the custodial singlet \emph{or} the neutral component of the fiveplet, we see it may be challenging to distinguish the GM and SGM models by simply measuring diphoton decays of $H_5$.
\begin{figure}[th]
\includegraphics[width=.45\textwidth]{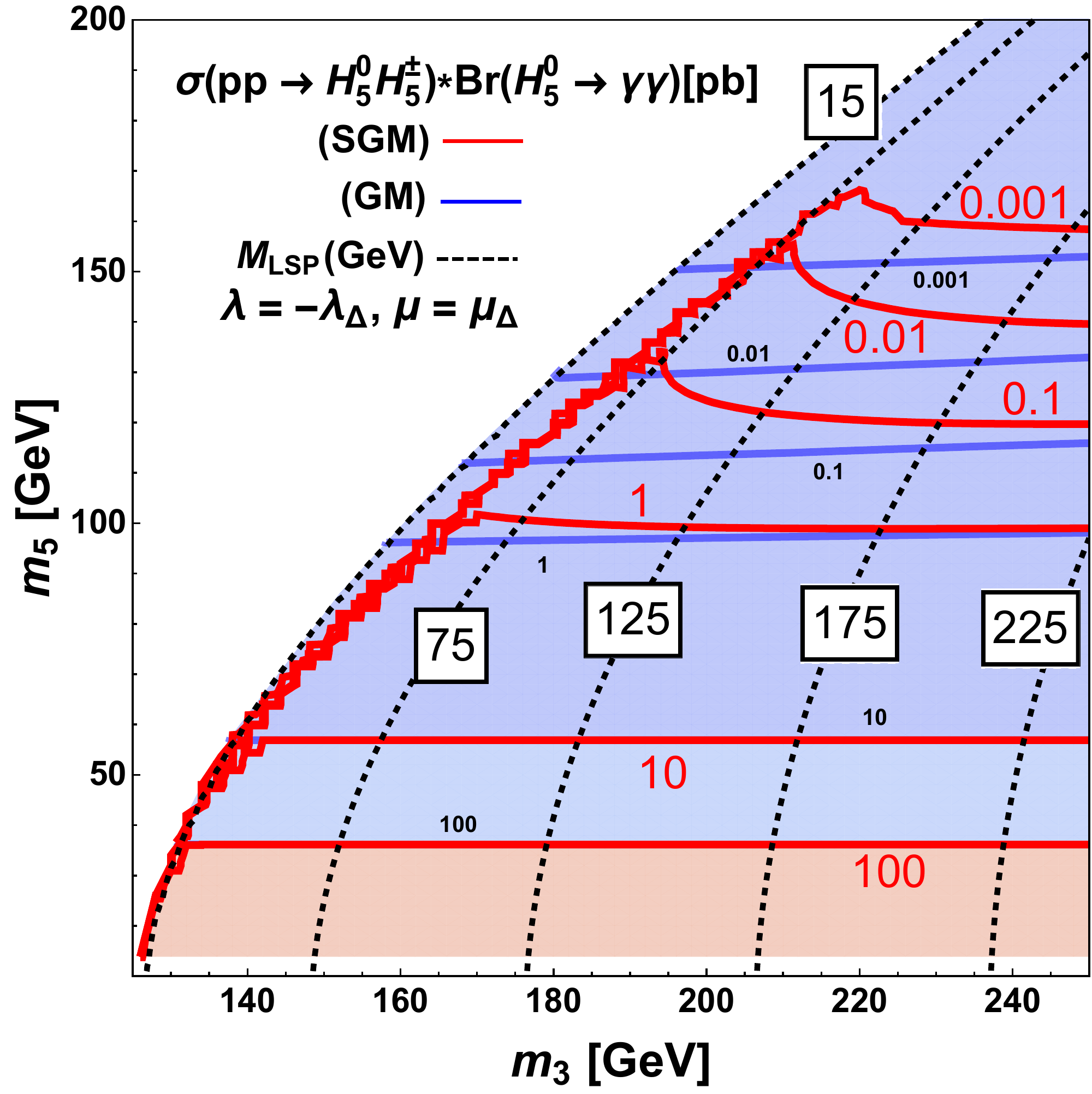}
\includegraphics[width=.45\textwidth]{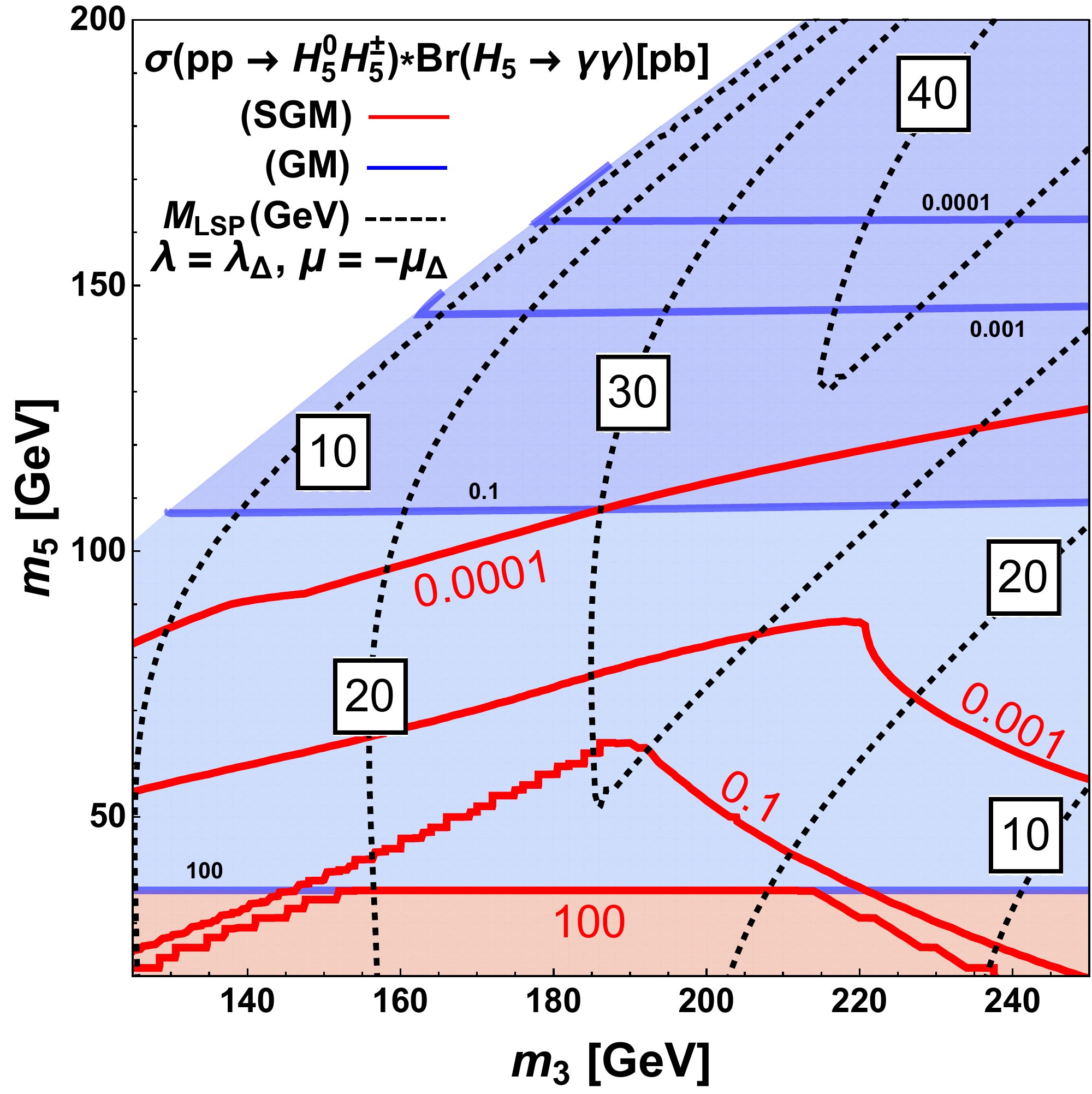}
\caption{{\bf Top:~}Contours of the custodial fiveplet (pair) production cross section ($pb$) times branching ratio into photons at a 13 TeV LHC.~Results for the GM model (blue solid lines with background color) and SGM model (red solid line) are shown with contours of the LSP mass (black dashed lines) and inputs defined for Point 1 in~\eref{points12}.~{\bf Bottom:~}Same as top, but for Point 2 in~\eref{points12}.}
\label{fig:H5aaconts}
\end{figure}

For point 2 (bottom) with $\lambda = \lambda_\Delta$, allowing for smaller $\mu_\Delta$, we have drastically different contours and a much lighter LSP $\lesssim 45$~GeV.~The LSP is now composed of the neutral component of the lightest custodial triplet, while its charged components contribute to the diphoton width.~Here we see clearly effects of the fermionic superparters which enter at both tree level \emph{and} 1-loop.~In this case, simply measuring the fiveplet branching ratio (or partial width) into photons would allow us to distinguish between the SGM and GM models.

We also see that at such light LSP masses, the large invisible decay width into the LSP suppresses the branching ratio into photons to be $\lesssim\mathcal{O}(10^{-3})$.~This would also allow to potentially avoid direct search constraints from diphoton searches even for masses below the SM-like Higgs mass (but above half the $Z$ mass) which, for small enough triplet VEVs, are still allowed~\cite{Delgado:2016arn,Brooijmans:2016vro}.~At the same time, utilizing the Drell-Yan Higgs pair production mechanism present in all models of extended Higgs sectors~\cite{Akeroyd:2003xi,Delgado:2016arn}, it may be possible to explain excesses in diphoton searches.~For instance the recently observed $\sim 3\sigma$ excess in the diphoton spectrum at $\sim 95$~GeV by the CMS collaboration~\cite{CMS:2017yta} may be explained by a custodial fiveplet Higgs, but we leave an exploration of this interesting possibility to ongoing work~\cite{followup}.~A number of recent studies have also examined various possibilities for explaining this excess~\cite{Mariotti:2017vtv,Crivellin:2017upt,Fox:2017uwr,Haisch:2017gql}.

In contrast, for the GM model where there is no decay to an LSP available, fiveplet masses in this range are ruled out by diphoton searches at the LHC~\cite{Delgado:2016arn,Brooijmans:2016vro}.~This is true largely independently of the triplet VEV assuming the loop induced coupling to diphotons is dominated by the $W$ boson, as often the case in the absence of large couplings in the scalar potential.~One could in principle cancel the $W$ boson loop contribution with charged scalar loops to reduce the diphoton branching ratio and evade these constraints or, as done for a number of Higgs triplet models~\cite{Bahrami:2015mwa,Tait:2016qbg,Lu:2016ucn} including the GM model~\cite{Pilkington:2017qam}, add additional particles which contain a sufficiently light DM candidate, thus opening up an invisible decay channel.

We show in~\fref{H5zzconts} the same as in~\fref{H5aaconts}, but now for the $ZZ$ branching ratio.~Again we see the same qualitative relationship between the SGM and GM model as found for the diphoton channel.~In particular as the LSP masses becomes large, it becomes more difficult to distinguish the SGM from the GM model.~Furthermore, we see that when the LSP is light, the branching ratio for the fiveplet into $ZZ$, which occurs at tree level for non-zero triplet VEV, can be highly suppressed as the decay width into the LSP becomes dominant.~In particular, we see that a highly suppressed branching ratio for custodial fiveplet decays into $ZZ$ pairs, even for sizable triplet VEV, may be a distinguishing feature of a GM-like model with supersymmetric origin.
\begin{figure}[tbh]
\includegraphics[width=.45\textwidth]{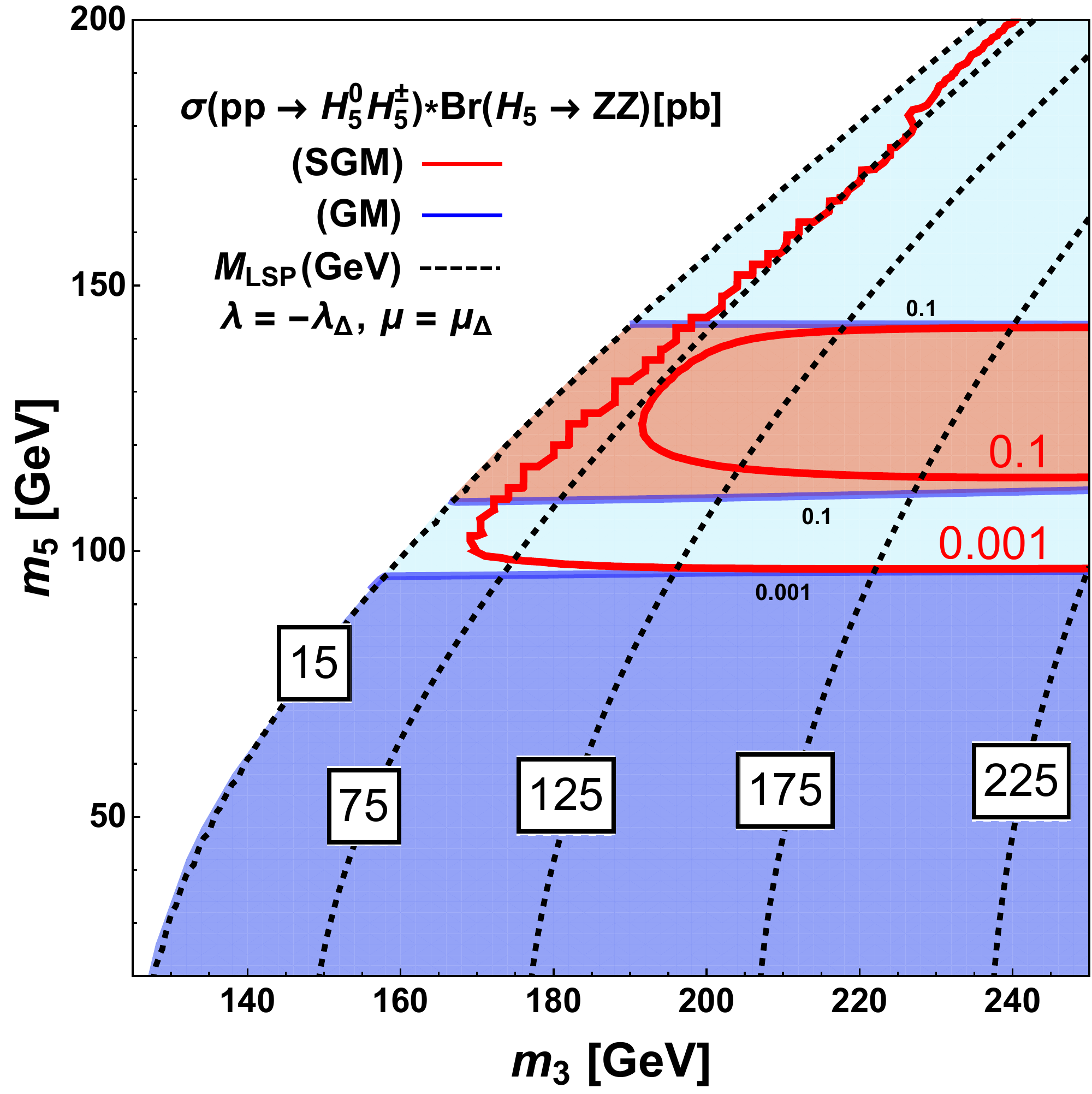}
\includegraphics[width=.45\textwidth]{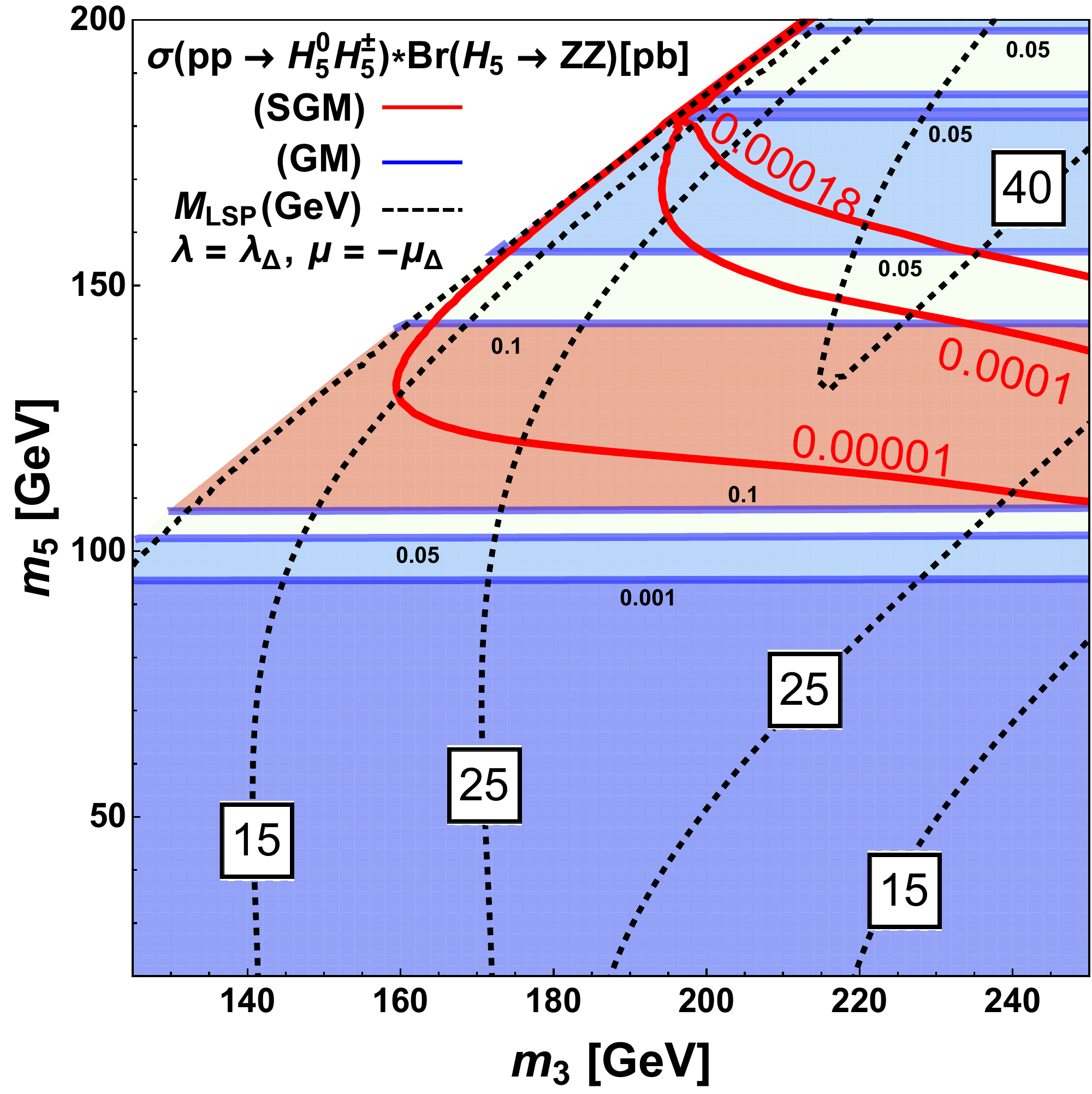}
\caption{{\bf Top:~}Contours of the custodial fiveplet (pair) production cross section ($pb$) times branching ratio into $ZZ$ pairs at a 13 TeV LHC.~Results for the GM model (blue solid lines with background color) and SGM model (red solid line) are shown with contours of the LSP mass (black dashed lines) and inputs defined for Point 1 in~\eref{points12}.~{\bf Bottom:~}Same as top, but for Point 2 in~\eref{points12}.}
\label{fig:H5zzconts}
\end{figure}

Finally, we have not examined the dependence on the triplet VEV in detail here since the focus of this study is on how the SGM arises from the SCTM and how it maps onto the GM model.~The dependance on the triplet VEV is the same in both the SGM and GM models so it does not serve as a useful probe for distinguishing them, but a detailed study including an electroweak precision analysis of the SGM model deserves future investigation.

\subsection{The Higgs golden ratio}\label{sec:ratio}

Next we examine decays of the SM-like Higgs boson into $\gamma\gamma$ and $ZZ$ pairs.~More precisely we examine a quantity dubbed the `Higgs golden ratio' defined as~\cite{Djouadi:2015aba},
\bea
\label{eq:goldenratio}
\mathcal{D}_{\gamma \gamma}^{SGM(GM)} &\equiv&
\frac{ Br^{GM (SGM)}_{h\to\gamma\gamma}/Br^{GM (SGM)}_{h\to ZZ} }
{ Br^{SM}_{h\to \gamma\gamma}/Br^{SM}_{h\to ZZ} }  .
\eea
where we have $\mathcal{D}_{\gamma \gamma}^{SM} \equiv 1$ using $Br_{h\to\gamma\gamma}^{SM}=0.228\%,\,Br_{h\to ZZ}^{SM}=2.64\%$ for the SM branching ratios~\cite{Agashe:2014kda}.~In addition to being precisely measured final states, this ratio benefits from the fact that many uncertainties coming from production effects cancel.~This should allow for measurements of $\mathcal{D}_{\gamma \gamma}$ to eventually reach $\mathcal{O}(1\%)$ precision~\cite{Djouadi:2015aba} at a high luminosity LHC.

In~\fref{Dhaaconts} we show contours of $\mathcal{D}_{\gamma \gamma}$ as a function of $m_5$ versus $m_3$ for a mass range $\sim 100 - 250$~GeV in both the GM (blue solid lines with background color) and SGM models (red solid line).~The black dashed lines indicate contours of the LSP mass in the SGM.~We again see a similar relationship between the SGM and GM model as found for the custodial fiveplet decays.~In particular, as the LSP masses are taken larger, the GM and SGM predictions for $\mathcal{D}_{\gamma \gamma}$ begin to converge while when the LSP is light there can be striking differences.~We also see that potentially observable deviations from $\mathcal{D}^{SM}_{\gamma \gamma} = 1$ may be possible at the LHC over a range of custodial fiveplet and triplet masses or conversely, that large regions of parameter space can be ruled out as measurements of $\mathcal{D}_{\gamma \gamma}$ become more precise.~We leave further investigation of this ratio as well differential studies in the $h\to 4\ell$ and $h\to 2\ell\gamma$ channels to future work.
\begin{figure}[tbh]
\includegraphics[width=.45\textwidth]{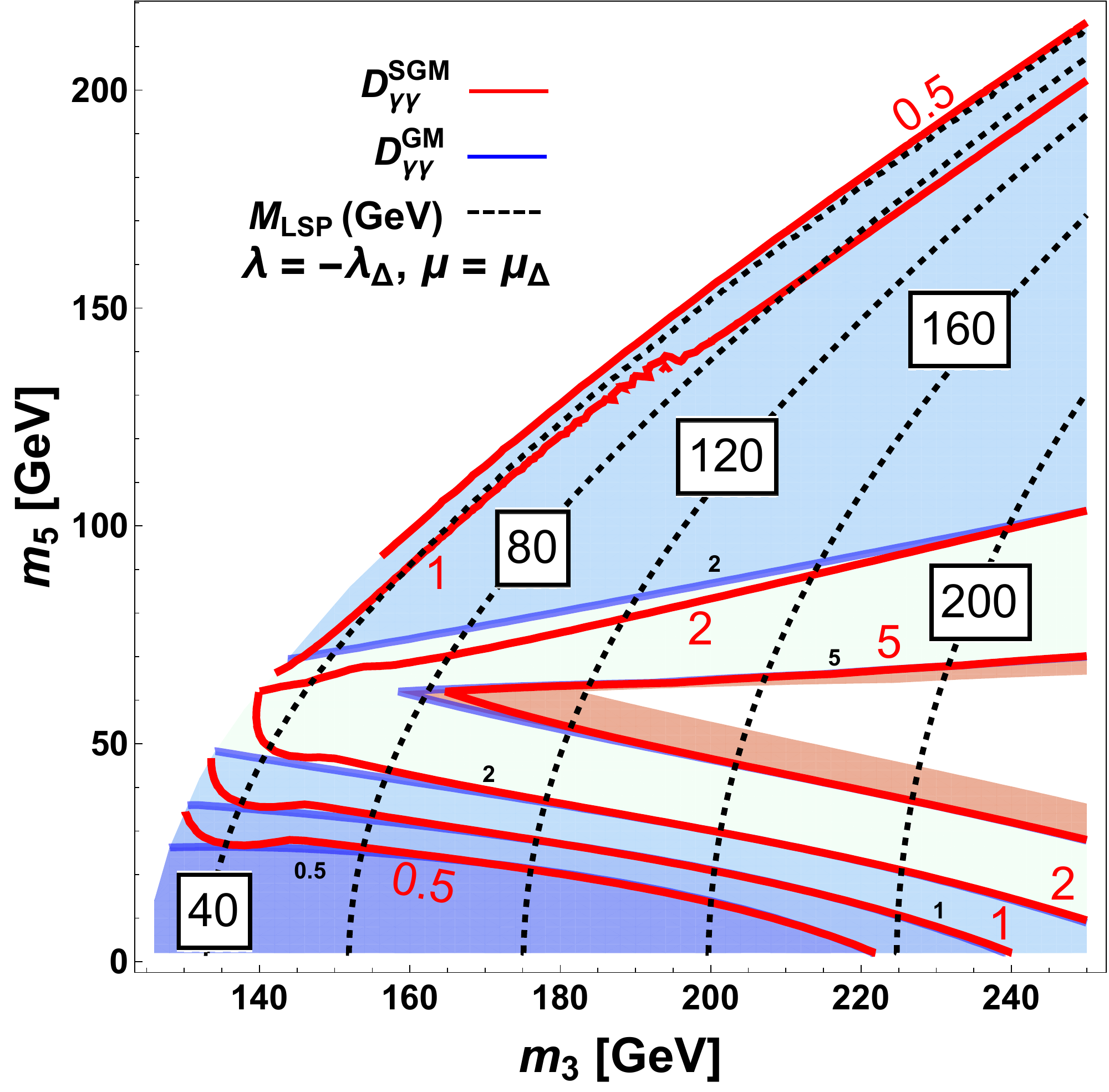}
\includegraphics[width=.45\textwidth]{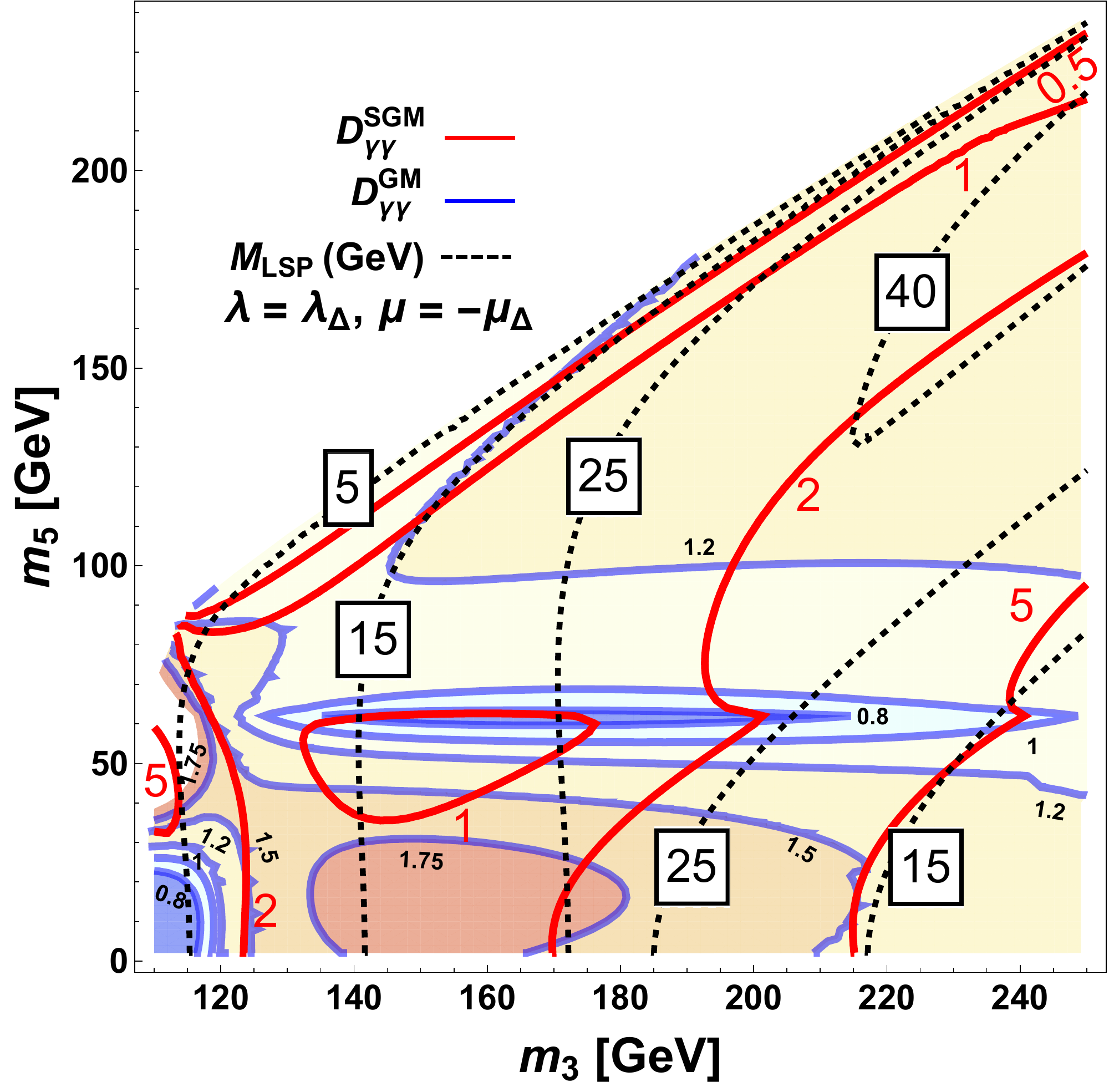}
\caption{{\bf Top:~}Contours of the `Higgs golden ratio' $\mathcal{D}^{SM}_{\gamma \gamma} $ for the SM-like Higgs as defined in~\eref{goldenratio} for the GM model (blue solid lines with background color) and SGM model (red solid line) with contours of the LSP mass (black dashed lines) and inputs defined for Point 1 in~\eref{points12}.~{\bf Bottom:~}Same as top, but for Point 2 in~\eref{points12}.}
\label{fig:Dhaaconts}
\end{figure}

\section{Summary and Conclusions}\label{sec:disc}

We have shown that the well known Georgi-Machacek (GM) model~\cite{Georgi:1985nv,Chanowitz:1985ug} can be realized as a limit of the recently constructed Supersymmetric Custodial Higgs Triplet Model (SCTM)~\cite{Cort:2013foa,Garcia-Pepin:2014yfa} and have dubbed this limit the Supersymmetric GM (SGM) model.~In doing so we have realized a weakly coupled origin for the GM model at the electroweak scale, in contrast to the more commonly envisioned strongly coupled composite Higgs scenarios.~A supersymmetric origin for the GM model comes with many theoretical and phenomenological benefits, including a possible dark matter candidate, which we have discussed.~As part of demonstrating this limit, we have also derived a mapping between the SGM and GM model which we use to show that a supersymmetric origin for the GM model implies constraints on the Higgs potential which are otherwise not be present in conventional GM model constructions.~We have also discussed the superpartner fermion sector and LSP of the SGM as well as derived the custodial fermion mass matrix.

We then discussed using diboson signals to distinguish the SGM from the GM model at the LHC focusing in particular on diphoton and $ZZ$ decays of the custodial fiveplet scalar and the SM-like Higgs boson.~We have studied a pair of benchmarks scenarios to demonstrate that along the `slices' of parameter space in the GM model where the supersymmetric constraints are satisfied (see~\eref{gmcon}), the GM-model can appear to be very similar to the SGM model depending on the mass scale of the fermion superpartner sector, which we characterize with the (neutralino) LSP mass.~In general we find that when the LSP is light in the SGM, there are striking differences between the SGM and GM model phenomenology while when the LSP is heavy, the two models can appear very similar.

We have also discussed the possibility that light diphoton signals in the GM and SGM models may be observable at the LHC, but leave a detailed study for ongoing work~\cite{followup}.~We have not performed a comprehensive parameter space scan, but results from previous studies suggest much of the parameter space considered here are still allowed by experiment~\cite{Aaltonen:2016fnw,Delgado:2016arn,Fox:2017uwr,Pilkington:2017qam}.~A detailed scan to establish the allowed parameter space including relevant experimental constraints as well as a more comprehensive study of signals is left to ongoing work.~An NLO analysis including custodial breaking effects may be important in some cases, but also left to future work.

The GM model has long been a phenomenologically interesting possibility for an extended Higgs sector with potentially striking LHC signals.~The SGM gives a supersymmetric possibility for this model with particular and potentially striking differences that can be searched for at the LHC.~We hope that this will encourage current LHC experiments, which have dedicated searches for signals of the GM model~\cite{deFlorian:2016spz}, to augment their analyses to also generally include signals of the SGM model. 

  \clearpage
\noindent
{\bf Acknowledgements:}~We thank Mariano Quiros and Mateo Garcia-Pepin for many useful discussions.~The work of R.V.M.~is supported by MINECO,~FPA 2016-78220-C3-1-P,~FPA 2013-47836-C3-2/3-P (including ERDF), and the Juan de la Cierva program,~as well as by Junta de Andalucia Project FQM-101.~The work of R.V.~is partially supported by the Sam Taylor fellowship.~K.X. is supported by U. S. Department of Energy under Grant No. DE-SC0010129.~R.V.M.~would like to thank the Mainz Institute for Theoretical Physics (MITP) for its hospitality and partial support during the completion of this work as well as Fermilab National Accelerator Laboratory and Northwestern University for their hospitality.~R.V.~would like to thank the University of Granada for their hospitality while part of this work was done.~K.X.~also thanks Fermilab for their hospitality and partial support during the completion of this work.

\section{Appendix}\label{sec:app}
To facilitate comparison, in this appendix we give the necessary details to derive the mapping in~\eref{relation} utilizing the Higgs fields and SCTM potential in a basis that is more commonly used in conventional GM model constructions~\cite{Gunion:1989ci,Gunion:1990dt,Hartling:2014zca}.~The electroweak doublet and triplet Higgs superfields in the SCTM~\cite{Cort:2013foa} can be arranged into $\bf{(2,\bar2)}$ and $\bf{(3,\bar3)}$ representations of the global $SU(2)_L\otimes SU(2)_R$ symmetry as,
\beq\label{eq:DH}
\Phi = 
\begin{pmatrix}
H_1^{0} & H_2^+\\
H_1^- & H_2^0
\end{pmatrix}  
,~
X = 
\begin{pmatrix}
\chi^{0} & \phi^+ & \psi^{++}\\
\chi^- & \phi^0 & \psi^+\\
\chi^{--}& \phi^-& \psi^0 \, 
\end{pmatrix}  .
\eeq
which transform as $\Phi \rightarrow U_L\Phi U_R^\dagger$ and $X \rightarrow U_L X U_R^\dagger$.

The manifestly $SU(2)_L\otimes SU(2)_R$ symmetric superpotential can then be written in terms of these fields as,
\beq
\begin{split}
W_0 & =
2 \lambda \,Tr[ \Phi_c \tau_i \Phi \tau_j ] [U X U^\dagger]_{ij} \\
& \quad -  \frac{\lambda_\Delta}{6} Tr[ X_c t_i X t_j] [U X  U^\dagger]_{ij} \\ 
&\quad -  \frac{\mu}{2} Tr[ \Phi_c \Phi ]  +
\frac{\mu_\Delta}{2} Tr[ X_c X] ,
\end{split}
\eeq 
where $\tau_i = \sigma_i/2$ and $t_i$ are the two and three dimensional
representations respectively of the $SU(2)$ generators as defined in~\cite{Hartling:2014zca}, along with the matrix $U$ which is used to rotate the matrix field $X$ into the Cartesian basis~\cite{Aoki:2007ah}.~We have also defined $X_c=C X^T C$ and $\Phi_c=\sigma_2 \Phi^T \sigma_2$ respectively, where the symmetric matrix $C$ is given by,
%
\beq
\label{eq:c2c3}
~\\
C
=
\begin{pmatrix}
0 & 0 & 1 \\
0 & -1 & 0 \\
1& 0&  0
\end{pmatrix}  .
\eeq\\
%
With these definitions,   $X_c$ and $\Phi_c$, have the same tranformation properties under the global $SU(2)_L\otimes SU(2)_R$ symmetry as  $X^\dagger$ and $\Phi^\dagger$ respectively.

The F-term contribution to the Higgs potential can then be obtained from the superpotential following the usual procedure or by listing all the possible $SU(2)_L\otimes SU(2)_R$ invariants of dimension four or less that can be built using the fields $(\Phi,\, \Phi_c,\,\Chi,\, \Chi_c)$ and their hermitian conjugates.~The result of this procedure gives,
\begin{widetext}
\beq
\label{eq:vsctm2} 
\begin{split}
V_{\rm{F}} & =
\mu ^2 \Tr{ \Phid \Phi}  +  \mu _{\Delta }^2  \Tr{ \Chid \Chi}  
+ \lambda ^2 \biggl( \Tr{ \Phid \Phi}^2 - \frac{1}{4} \Tr{ \Phic\Phi}\, \Tr{ \Phicstar\Phid}  
+  \txcIc \\
& \quad +  \txcIf  -  \txcIh   - \txcIcO  \biggr) \\
& \quad + \dfrac{\lambda \mu _{\Delta }}{2} \left(\txx \right)  -  \dfrac{\lambda _3^2}{2}  \left(\txcO \right)  \\
& \quad -  \dfrac{\lambda _3 \lambda}{4}  \left( \txcTc \right) 
 -  \dfrac{\lambda _3 \mu _{\Delta }}{2} \left( \txcEc  \right)  \\
& \quad -  \lambda  \mu  \left( \txcFc  \right) ,
\end{split} 
\eeq
\end{widetext}
where the notation, relative signs, and numerical factors are arranged so that~\eref{vsctm2} agrees precisely with~\eref{vsctm} when both are written in component form.~The soft supersymmetry breaking potential is given by,
\begin{widetext}
\bea
\label{eq:vsoft2}
  V_{\rm{soft}} &=&
   m_H^2\, \Tr{ \Phi^\dagger \Phi }  + m_\Delta^2\, \Tr{ X^\dagger X } 
 +  \biggl( \dfrac{B_\Delta}{2} \, \Tr{ X_c X } - \dfrac{B}{2} \, \Tr{ \Phi_c \Phi }  \nn
&+&  \dfrac{A_\lambda}{2}\, \Tr{ \Phi_c  \sigma_i \Phi \sigma_j } (U X U^\dagger)_{ij}
- \dfrac{A_\Delta}{6}\,  \Tr{ X_c t_i X t_j } (U X U^\dagger)_{ij}  + c.c. \biggr), 
\eea
\end{widetext}
where again the signs and conventions are chosen so that~\eref{vsoft2} matches~\eref{vsoft} exactly when written in component form.

With these conventions the `un-complexification' constraint in~\eref{DelHmap} takes the following form,    
\beq
\label{eq:DelHmap2}
X_c = X^\dagger, \quad \Phi_c = \Phi^\dagger\, .
\eeq
Imposing these conditions\,\footnote{Note these conditions imply the substitutions on the component electroweak fields in~\eref{DH} of the form:
$\psi^{0} \rightarrow \chi^{o*},~
\psi^{-} \rightarrow \chi^{-},~
\psi^{--} \rightarrow \chi^{--},~ 
H_2^o \rightarrow H_1^{o*},~
H_2^{+} \rightarrow H_1^{+}$, which also fixes the phase conventions $H_1^{-\ast} = -H_1^+$ and $\chi^{-\ast} = -\chi^+$.}
on~\eref{DH} leads to the GM model fields as defined in~\cite{Hartling:2014zca}.~Furthermore, as found in~\sref{GMmodel}, after imposing these constraints, the expression for the potential $V_{\rm{F}} + V_{\rm{soft}}$ given by~\eref{vsctm2} plus~\eref{vsoft2} can again be mapped to\,\footnote{Again the D-terms reduce to zero when imposing~\eref{DelHmap2}.}
a Higgs potential of the same form as found in the GM model~\cite{Hartling:2014zca},
\begin{widetext}
\bea
\label{eq:vgm2} 
V_{GM} &=& 
\frac{\mu_2^2}{2} \Tr{ {\Phi}^\dagger {\Phi} } 
+ \frac{\mu_3^2}{2} \Tr{ {X}^\dagger {X} }
+ \lambda_1 \Tr{ {\Phi}^\dagger {\Phi}  }^2 
+ \lambda_2 \Tr{ {\Phi}^\dagger {\Phi} }\, \Tr{ {X}^\dagger {X} }
+ \lambda_3 \Tr{ {X}^\dagger X {X}^\dagger {X} } 
+  \lambda_4  \Tr{ {X}^\dagger {X} }^2 \nn
&-& \lambda_5 \Tr{ {\Phi}^\dagger \tau^a  {\Phi}  \tau^b } 
\Tr{ {X}^\dagger t^a {X} t^b  } 
- M_1 \Tr{ {\Phi}^\dagger \tau^a {\Phi}  \tau^b } (U {{X}} U^\dagger)_{ab} 
-  M_2 \Tr{ {X}^\dagger t^a  {X} t^b } (U {\bar{X}} U^\dagger)_{ab} . \nonumber
\eea
\end{widetext}
After matching coefficients of common operators  in $V_{\rm{F}} + V_{\rm{soft}}$ (after applying~\eref{DelHmap2}) and $V_{GM}$ we again arrive at the same mapping between scalar potential parameters as given in~\eref{relation}.~Note that the second and third terms of the second line in~\eref{vsctm2} reduce to zero upon imposing~\eref{DelHmap2} and do not appear in~\eref{vgm2}.
%



\bibliographystyle{apsrev}
\bibliography{refs_SGMv2}

\end{document}